\documentclass[aps,prb,twocolumn,groupedaddress,showpacs,%
               nobalancelastpage,floatfix]{revtex4}
\usepackage{amsmath,amssymb,amsfonts}
\usepackage{graphicx}
\begin{document}
\title{Diagrammatic approximations for the 2d quantum antiferromagnet:
       exact projection of auxiliary fermions}
\author{Jan Brinckmann}
\author{Peter W{\"o}lfle}
\affiliation{Institut f{\"u}r Theorie der Kondensierten Materie,
             Universit{\"a}t Karlsruhe, 
             D-76128 Karlsruhe, Germany}
\date{22 March 2008}
\begin{abstract}
  We present diagrammatic approximations to the spin dynamics of the
  2d Heisenberg antiferromagnet for all temperatures, employing an
  auxiliary-fermion representation. The projection onto the physical
  subspace is effected by introducing an imaginary-valued chemical
  potential as proposed by Popov and Fedotov. The method requires that
  the fermion number at any lattice site is strictly conserved.  We
  compare results obtained within a self-consistent approximation
  using two different auxiliary-particle projection schemes, (1) exact
  and (2) on average.  Significant differences between the two are
  found at higher temperatures, whereas in the limit of zero
  temperature (approaching the magnetically ordered ground state)
  identical results emerge from (1) and (2), providing the
  qualitatively correct dynamical scaling behavior. An interpretation
  of these findings is given. We also present in some detail the
  derivation of the approximation, which goes far beyond mean-field
  theory and is formulated in terms of complex-valued spectral
  functions of auxiliary fermions.
\end{abstract}
\maketitle
\def\ppv{{\textrm{ppv}}}
\def\ave{{\textrm{av}}}
\def\av{{\textrm{av}}}
\def\ff{{\widetilde{H}}}

\section{Introduction and review}
\label{sec-intro}
Auxiliary particles are a widely used tool in the theory of correlated
electron systems. The principal difficulty in the treatment of these
systems is the strong Coulomb repulsion $U$ for two electrons on the
same localized orbital, usually of $d$ or $f$ character. In effective
model Hamiltonians like the Heisenberg, $t$-$J$\,, or Kondo model the
large $U$ leads to a Gutzwiller projection onto the quantum-mechanical
subspace, where none of the $d$- or $f$-orbitals may contain more than
one electron at a time.

In this paper we will focus on the spin-1/2 quantum Heisenberg
antiferromagnet in two spatial dimensions (2D) on a square
lattice\cite{man91},
\begin{equation}
  \label{eqn-hbergham}
  H= \sum_{<i, j>}J\,\mathbf{S}_i \mathbf{S}_j
     \;\;,\;\;\;
     J> 0
\end{equation}
The sum covers all nearest-neighbor pairs \mbox{$<i, j>$}\,. The model
(\ref{eqn-hbergham}) may be obtained from the single-band Hubbard
model for large $U$ with nearest-neighbor hopping amplitude $t$
(leading to $J= 2t^2/U$) in the limit of a half-filled band with one
electron per site\cite{grosjoyntrice87}. It represents the simplest
low-energy model for two-dimensional Mott insulators, in particular
the CuO-planes in the undoped parent compounds for high-temperature
superconductors\cite{hay90,keim92,kim01}.

The restriction on states with no doubly occupied sites is reflected
in the non-canonical commutation relations of spin operators,
\begin{equation}
  \label{eqn-comm}
  {}[\,S^x\,, S^y\,{}]
    = i S^z \ne \textrm{c-number}.
\end{equation}
For an analytical approach to the model, Eq.(\ref{eqn-comm}) poses a
severe difficulty, since the standard Wick's theorem and many-body
techniques cannot be applied\cite{negorl,grekei81,bic87,russenbuch}. A
convenient way of circumventing this difficulty is to represent the
spin operators in terms of canonical auxiliary-particle operators, of
either fermionic\cite{abr65} or bosonic\cite{aaa88} character. The
cost of this concept is the extension of the Hilbert space into
unphysical sectors. These unphysical states have to be removed by
imposing a constraint. In this work we use auxiliary fermions,
\begin{equation}
  \label{eqn-auxferm}
  S^\mu_i
    = \frac{1}{2}
      \sum_{\alpha, \bar{\alpha}}
      f^\dagger_{i\alpha}
      (\sigma^\mu)^{\alpha \bar{\alpha}}
      f_{i\bar{\alpha}}
      \;\;,\;\;\;
  Q_i= \sum_{\alpha}
       f^\dagger_{i\alpha} f_{i\alpha}
     = 1 \;\;,
\end{equation}
$\sigma^\mu$\,, $\mu= x, y, z$ are the Pauli matrices, and $\alpha=
\uparrow, \downarrow$ is the fermion-spin index. Here and in the
following we let $\hbar\equiv 1$\,. The representation
(\ref{eqn-auxferm}) fulfills the commutation relations
(\ref{eqn-comm})\,. The Fock space of the auxiliary fermions
$f_{i\alpha}$ is spanned by the states
\begin{subequations}
  \label{eqn-states}
\begin{eqnarray}
  \label{eqn-physstates}
    \textrm{physical:} & \ &
      |\uparrow\rangle= f^\dagger_{i\uparrow}|0\rangle  \;\;,\;\;\;
      |\downarrow\rangle= f^\dagger_{i\downarrow}|0\rangle
      \\
  \label{eqn-unphysstates}
    \textrm{\emph{un}physical:} & \ &
      |0\rangle \;\;,\;\;\;
      |2\rangle= f^\dagger_{i\uparrow} f^\dagger_{i\downarrow}|0\rangle
      \\ \nonumber
\end{eqnarray}
\end{subequations}
where $|0\rangle$ denotes the vacuum, $f_{i\alpha}|0\rangle\equiv
0$\,.

The projection onto the physical subspace can be performed in several
ways\cite{bic87}, and for impurity models like the Kondo model or the
single-impurity Anderson model the projection of auxiliary particles
is a standard technique\cite{abrmig70,biccoxwil87}. (For the latter
model, auxiliary bosons have to be added to the scheme\cite{col84}.)
However, in lattice models like Eq.(\ref{eqn-hbergham}) the constraint
$Q_i= 1$ has to be enforced on each site $i$ independently. This leads
to the so-called excluded-volume problem\cite{bro61,grekei81} and
prohibits an infinite-order resummation of the perturbation series in
$J$\,. In the limit of high spatial dimension the problem can be
evaded within the (extended) dynamical mean-field theory. Here the
infinite lattice system is approximately mapped onto a single
site\cite{kur85,metvol89,sismith96} or a small cluster\cite{note-dca}
coupled to a bath.

An alternative path starts from a mean-field-like treatment of the
constraint, where the auxiliary charge $Q_i= 1$ is fixed merely on the
thermal average\cite{newrea87},
\begin{equation}
  \label{eqn-meancons}
  Q_i \to \langle Q_i\rangle
    = \langle Q_1\rangle
    = \sum_{\alpha}\langle f^\dagger_{1\alpha} f_{1\alpha}\rangle
    = 1\;.
\end{equation}
This condition is introduced into the Hamiltonian
(\ref{eqn-hbergham}), (\ref{eqn-auxferm}) through a chemical potential
$\mu^f$ for the auxiliary fermions. Due to the particle--hole symmetry
of (\ref{eqn-hbergham}) we have $\mu^f= 0$\,. The approximation
(\ref{eqn-meancons}) is of great advantage, since now the perturbation
theory in $J$ starts from non-interacting fermions, and we can make
use of the standard Feynman-diagram techniques.
Eq.(\ref{eqn-meancons}) is also the starting point for numerous
mean-field theories of correlated electron systems. An improvement of
the mean-field-like constraint (\ref{eqn-meancons}) in a perturbative
fashion has frequently been made by generalizing $\mu^f$ to a
fluctuating Lagrange multiplier (see, e.g., Ref.\
\onlinecite{bic87})\,.

\emph{The Popov--Fedotov approach:} The method proposed by Popov and
Fedotov\cite{popfed88} enables one to enforce the auxiliary-particle
constraint \emph{exactly} within an analytical calculation for the
infinite system\cite{grosjohn90,boukis99,kisopp00,dilric06}. The
approach starts from a ``grand-canonical ensemble'',
\begin{equation}
  \label{eqn-grand}
  H\to H^\ppv=
    H - i\frac{\pi}{2}T\sum_{i}Q_i  \;\;,\;\;\;
  Q_i= \sum_{\alpha}f^\dagger_{i\alpha} f_{i\alpha}  \;\;,\;\;\;
\end{equation}
with a homogeneous, \emph{imaginary-valued} chemical potential
\begin{equation}
  \label{eqn-ppv}
  \mu^f= i\frac{\pi}{2}k_BT\equiv i\frac{\pi}{2}T\;.
\end{equation}
$H$ is the spin Hamiltonian (\ref{eqn-hbergham}), written in terms of
the auxiliary-fermion operators (\ref{eqn-auxferm})\,.

There are two main requirements for the Popov--Fedotov method to work:
The first is the conservation of the auxiliary charge $Q_i$ on each
lattice site,
\begin{equation}
  \label{eqn-qcons}
  {}[\,Q_i\,, H\,{}]= 0 \;\;,\;\;\; 
  i= 1, 2, \ldots, N_L \;\;,
\end{equation}
where $N_L$ denotes the number of sites (i.e., spins) in the system
(\ref{eqn-hbergham})\,. Since also ${}[\,Q_i\,, Q_j\,{}]= 0$\,, the
eigenstates of $H$ and $H^\ppv$ in the Fock space of the fermions can
be specified by some auxiliary-charge configuration
\begin{equation}
  \label{eqn-config}
  c_Q= (Q_1, Q_2, \ldots, Q_{N_L}) \;\;,\;\;\;
  Q_i\in \{0, 1, 2\}\;.
\end{equation}
Physical states belong to the subspace with the configuration
\begin{equation}
  \label{eqn-physconfig}
  c_Q^{phys}= (1, 1, \ldots, 1)\;.
\end{equation}
For a given configuration $c_Q$ the Hamiltonian (\ref{eqn-hbergham})
has a specific set of eigenstates with quantum numbers $n_Q$ and
energies $E(c_Q, n_Q)$\,, i.e., Schr{\"o}dinger's equation reads
\begin{equation}
  \label{eqn-sgl}
  H|c_Q, n_Q\rangle
    = E(c_Q, n_Q)|c_Q, n_Q\rangle\;.
\end{equation}
Consider now the partition function $Z^\ppv$ for the
grand-canonical Hamiltonian (\ref{eqn-grand}),
\begin{eqnarray}
  Z^\ppv
    & = &  \label{eqn-zppdef}
      \textrm{Tr}^f{}[\,e^{-\beta H^\ppv}\,{}]
      \\  \nonumber
    & = & \sum_{c_Q}\sum_{n_Q}
      \langle c_Q, n_Q|\,
        e^{-\beta H^\ppv}\,
        |c_Q, n_Q\rangle\;\;,
\end{eqnarray}
$\textrm{Tr}^f$ denotes the trace in the enlarged Hilbert space
of the auxiliary fermions, and $\beta= 1/k_B T\equiv 1/T$\,. With
Eqs.(\ref{eqn-sgl}) and (\ref{eqn-grand}) it becomes
\begin{eqnarray*}
  Z^\ppv
    & = &  \sum_{Q_1, \ldots, Q_{N_L}= 0}^2\;
      \sum_{n_Q}\cdot
      \\
    &   & \mbox{}\cdot
      \langle c_Q, n_Q|\,
      e^{-\beta E(c_Q, n_Q)}\,
      e^{i\frac{\pi}{2}Q_1}\cdots
      e^{i\frac{\pi}{2}Q_{N_L}}\,
      |c_Q, n_Q\rangle
\end{eqnarray*}

In addition to the $Q_i$-conservation, the Popov--Fedotov method
requires that the Hamiltonian and the operators appearing in physical
(i.e., observable) correlation functions destruct the unphysical
states $|0\rangle$\,, $|2\rangle$\,. In the present case, Hamiltonian
and correlation functions are composites of spin operators, and we
have, using Eqs.(\ref{eqn-auxferm}) and (\ref{eqn-unphysstates}),
\begin{equation}
  \label{eqn-destruct}
  S_i^\mu|0\rangle= 0  \;\;,\;\;\;
  S_i^\mu|2\rangle= 0\;.
\end{equation}
Consider an arbitrary site $l$ with an unphysical auxiliary charge
$Q_l\ne 1$\,. From Eq.(\ref{eqn-destruct}) it follows
\begin{equation}
  \label{eqn-equalen}
  Q_l\ne 1\,: \quad
  \left.E(c_Q, n_Q)\right|_{Q_l= 0}
    = \left.E(c_Q, n_Q)\right|_{Q_l= 2}\;\;,
\end{equation}
that is, the spin at site $l$ seems to be removed from the Hamiltonian
for all states $|c_Q, n_Q\rangle$ with $Q_l= 0$ or $2$\,. Accordingly,
the contribution from $Q_l= 0, 2$ to $Z^\ppv$ is proportional to
\begin{equation}
  \label{eqn-cancel}
  \sum_{Q_l= 0, 2}\,e^{i\frac{\pi}{2}Q_l}
    = (1 + e^{i\pi})
    = 0\;.
\end{equation}
In that way, the unphysical contributions from all sites $l= 1,
\ldots, N_L$ cancel in the grand-canonical partition function, i.e.,
only the physical charge configuration $c_Q^{phys}$\,,
Eq.(\ref{eqn-physconfig}), survives in $Z^\ppv$\,:
\begin{displaymath}
  Z^\ppv
    = (i)^{N_L}
      \sum_n
      \langle n|\,
        e^{-\beta E_n}\,
        | n\rangle\;.
\end{displaymath}
Here $|n\rangle$ and $E_n$ denote the eigenstates and -energies of the
Hamiltonian (\ref{eqn-hbergham}) in the physical subspace,
\begin{equation}
  \label{eqn-physen}
  \begin{array}[c]{c}
    \displaystyle
    |n\rangle= |c_Q^{phys}, n_Q\rangle  \;\;,\;\;\;
    E_n= E(c_Q^{phys}, n_Q)\;\;,
      \\[2ex]  \displaystyle 
    H|n\rangle = E_n|n\rangle\;.
  \end{array}
\end{equation}
Thus we end up with
\begin{equation}
  \label{eqn-physpart}
  Z^\ppv = (i)^{N_L}\,Z\;\;,
\end{equation}
i.e., up to a constant prefactor, the (canonical, physical) partition
function $Z$ for the Heisenberg model (\ref{eqn-hbergham}) is given by
$Z^\ppv$\,.

The above argument, originally presented by Popov and Fedotov in Ref.\
\onlinecite{popfed88}\,, is extended to Green's functions in Appendix
\ref{sec-app-greens}\,. It is found that any correlation function of
spin operators may be calculated from the grand-canonical Hamiltonian
(\ref{eqn-grand}).  In particular, the imaginary time-ordered spin
susceptibility
\begin{equation}
  \label{eqn-sus-def}
  \chi^{\mu\bar{\mu}}_{ij}(\tau - \tau')
    = \frac{1}{Z}
      \mbox{Tr}{}[\,
        e^{-\beta H}
        \mathcal{T}_\tau\{S^\mu_i(\tau)
        S^{\bar{\mu}}_j(\tau')\}\,{}]
\end{equation}
can be obtained from
\begin{equation}
  \label{eqn-sus-ppv}
  \chi^{\mu\bar{\mu}}_{ij}(\tau - \tau')
    = \langle \mathcal{T}_\tau\{S^\mu_i(\tau)
        S^{\bar{\mu}}_j(\tau')\}\rangle^\ppv
\end{equation}
with $\mu, \bar{\mu}\in \{x, y, z\}$\,. The expectation value is
calculated in the enlarged Fock space,
\begin{equation}
  \label{eqn-expect-ppv}
  \langle \ldots\rangle^\ppv
    = \frac{1}{Z^\ppv}
      \textrm{Tr}^f{}[\,e^{-\beta H^\ppv}\,\ldots\,{}]\;\;,
\end{equation}
with $Z^\ppv$ as defined in Eq.(\ref{eqn-zppdef}) above. The
``modified Heisenberg'' picture\cite{negorl} for spin operators reads
\begin{equation}
  \label{eqn-taudep}
  S^\mu_i(\tau)
    = e^{\tau H}S^\mu_i e^{-\tau H}
    = e^{\tau H^\ppv}S^\mu_i e^{-\tau H^\ppv}\;\;,
\end{equation}
using again Eq.(\ref{eqn-destruct})\,. Similarly, the local
magnetization is given by
\begin{equation}
  \label{eqn-magn-ppv}
  \langle S^\mu_i\rangle
    = \frac{1}{Z}\textrm{Tr}{}[\,
      e^{-\beta H} S^\mu_i\,{}]
    = \langle S^\mu_i\rangle^\ppv\;.
\end{equation}
It should be emphasized that expectation values of \emph{un}physical
operators are meaningless within the Popov--Fedotov scheme: E.g., the
auxiliary-fermion charge $Q_i$ introduced in Eq.(\ref{eqn-auxferm})
does not destruct the unphysical states (\ref{eqn-unphysstates}), and
one has
\begin{equation}
  \label{eqn-qexpect-ppv}
  \langle Q_i\rangle
    \boldsymbol{\ne} \langle Q_i\rangle^\ppv\;.
\end{equation}
For the l.h.s.\ we know that $Q_i= 1$ in the physical Hilbert space.
For the r.h.s., however, we obtain $\langle Q_i\rangle^\ppv= (1 +
i)$\,. The calculation is given in Appendix \ref{sec-app-greens}\,.

\emph{Average projection:} For comparison, we also want to use the
mean-field-like treatment of the auxiliary-fermion constraint
mentioned below Eq.(\ref{eqn-meancons})\,. Since the (real-valued)
chemical potential added to the Hamiltonian (\ref{eqn-hbergham}),
(\ref{eqn-auxferm}) turns out to be zero, due to the particle--hole
symmetry of (\ref{eqn-hbergham}), the calculation of a
spin-correlation function or magnetization amounts to just enlarging
the Hilbert space into the Fock space of the auxiliary fermions.  The
equivalents of the Eqs.(\ref{eqn-sus-ppv}), (\ref{eqn-magn-ppv}), and
(\ref{eqn-qexpect-ppv}) then read
\begin{equation}
  \label{eqn-sus-av}
  \chi^{\mu\bar{\mu}}_{ij}(\tau - \tau')
    \simeq \langle \mathcal{T}_\tau\{S^\mu_i(\tau)
        S^{\bar{\mu}}_j(\tau')\}\rangle^\ave
\end{equation}
and
\begin{equation}
  \label{eqn-magn-av}
  \langle S^\mu_i\rangle
    \simeq
    \langle S^\mu_i\rangle^\ave\;\;,
\end{equation}
whereas
\begin{equation}
  \label{eqn-qexpect-av}
  \langle Q_i\rangle
    = \langle Q_i\rangle^\ave
    = 1
\end{equation}
with
\begin{equation}
  \label{eqn-expect-av}
  \langle\ldots\rangle^\av
    = \frac{1}{Z^\av}
      \textrm{Tr}^f{}[\,
        e^{-\beta H}\ldots\,{}]
    \;\;,\;\;\;
  Z^\av
    = \textrm{Tr}^f{}[\,e^{-\beta H}\,{}]\;.
\end{equation}
The $\simeq$ sign in (\ref{eqn-sus-av}), (\ref{eqn-magn-av}) stands
for the error introduced by the uncontrolled fluctuations of the
fermion occupation numbers $Q_i$ into unphysical states. These
fluctuations are absent in the Popov--Fedotov scheme.

In the following sections, results from the exact and the averaged
constraint will be obtained within the same diagrammatic
approximations. The effect of the constraint on physical quantities
like the dynamical structure factor and magnetic transition
temperature is going to be studied. It will turn out that at
sufficiently low temperature the results for averaged and exactly
treated constraint become equal. In addition we will show in some
detail how a self-consistent approximation that goes far beyond
mean-field theory, can be worked out within the Popov--Fedotov
approach.

\section{Effect of the constraint: simple approximations}
\label{sec-simple}
In order to compare results for the Heisenberg model
(\ref{eqn-hbergham}) from average projection, Eq.(\ref{eqn-meancons}),
and exact projection using the Popov--Fedotov scheme,
Eq.(\ref{eqn-grand}), we consider a more general grand-canonical
Hamiltonian of auxiliary fermions, $\widetilde{H}= H - \mu^f\sum_i
Q_i$\,. With the model Hamiltonian (\ref{eqn-hbergham}) written
according to (\ref{eqn-auxferm}), it reads
\begin{eqnarray}
  \label{eqn-allgrand}
  \widetilde{H}
    & = &  -\mu^f\sum_{i}
      f^\dagger_{i\alpha} f_{i\alpha} \;+\;
      \\ \nonumber
    &   & \mbox{} +\;
      \frac{1}{2}\sum_{i,j}J_{ij}
      \frac{1}{4}
      \boldsymbol{\sigma}^{\alpha\bar{\alpha}}
      \boldsymbol{\sigma}^{\beta\bar{\beta}}
      f^\dagger_{i\alpha} f^\dagger_{j\beta}
      f_{j\bar{\beta}} f_{i\bar{\alpha}}\;.
\end{eqnarray}
Sums over spin indices $\alpha, \bar{\alpha}, \beta, \bar{\beta}$ are
implied. The antiferromagnetic coupling $J_{ij}$ is nonzero and equal
to $J>0$\,, if $i,j$ are nearest neighbors. Depending on the
projection method, the chemical potential $\mu^f$ takes the value
\begin{equation}
  \label{eqn-chem}
  \begin{array}[c]{lcrcl}
    \textrm{average projection:} & &
      \mu^f & = & 0\;\;,  \\[1ex]
    \textrm{exact projection:} & &
      \mu^f & = & \displaystyle
      i\frac{\pi}{2}T\;.
  \end{array}
\end{equation}
In the case of exact projection, the Hamiltonian (\ref{eqn-allgrand})
is no longer Hermitian. Nevertheless, physical quantities like the
dynamical structure factor for spin excitations or the magnetization
will come out real-valued.

Eq.(\ref{eqn-allgrand}) represents a system of canonical fermions with
a two-particle interaction $\sim J$ and may therefore be treated using
standard Feynman-diagram techniques\cite{negorl}. The bare
fermion-Green's function, written as a matrix in spin space, reads
\begin{equation}
  \label{eqn-baregf}
  \overline{G}^0_{ij}(i\omega)
    = \delta_{ij}\frac{1}{i\omega + \mu^f}\;\;,
\end{equation}
where $\omega= (2n + 1)\pi T$\,, $n\in\mathbb{Z}$ is a fermionic
Matsubara frequency. For the case of exact projection, $\mu^f= i\pi
T/2$\,, there is some common practice\cite{popfed88,boukis99,kisopp00}
to absorb $\mu^f$ into $i\omega$ and to re-define $\omega$
accordingly. Here we do not follow this line, but keep $\omega$ as
introduced above (i.e., fermionic).

\emph{Free spins ($J= 0$):} The simplest case is given by setting $J=
0$\,.  Using Eq.(\ref{eqn-auxferm}), the susceptibility
(\ref{eqn-sus-def}), to be calculated from Eq.(\ref{eqn-sus-ppv}) or
(\ref{eqn-sus-av}), is then given by a simple bubble,
\begin{eqnarray*}
  \chi_{ij}^{\mu\bar{\mu}}(i\nu)
    & = & -\frac{1}{4}T\sum_{i\omega}
      \textrm{Tr}^\sigma{}[\,\sigma^\mu
      \overline{G}^0_{ij}(i\omega + i\nu)\,\sigma^{\bar{\mu}}\,
      \overline{G}^0_{ji}(i\omega)\,{}]
      \\
    & = &  \delta_{i,j}\,\delta_{\mu,\bar{\mu}}\,\delta_{\nu, 0}\,
      \chi^0
\end{eqnarray*}
with, using Eq.(\ref{eqn-baregf}),
\begin{displaymath}
  \chi^0
    = \frac{1}{2T}f(-\mu^f) f(\mu^f)\;.
\end{displaymath}
$\nu= 2n\pi T$ is a bosonic Matsubara frequency, $\textrm{Tr}^\sigma$
denotes a trace in spin space, and $f(x)= 1/(e^{x/T} + 1)$ is the
Fermi function. Depending on the constraint method, the result is
\begin{equation}
  \label{eqn-chinul}
  \begin{array}[c]{lclcl}
    \textrm{average:} & & \displaystyle
      f(0)= \frac{1}{2} & \;\Rightarrow\; & \displaystyle
      \chi^0= \frac{1}{8T}\;\;,
      \\[2ex]
    \textrm{exact:} & & \displaystyle
      f(\pm i\frac{\pi}{2}T)
        = \frac{1}{1 \pm i} & \;\Rightarrow\; & \displaystyle
      \chi^0= \frac{1}{4T}\;.
  \end{array}
\end{equation}
Two observations are in order: The imaginary chemical potential
cancels out in the physical quantity $\chi^0$\,, and the result with
and without use of the exact auxiliary-particle projection is
qualitatively the same (Curie law). With merely average projection in
effect ($\mu^f= 0$), the spin moment $S(S + 1)$ extracted from the
Curie law $\chi^0(T)\propto S(S + 1)/T$ is reduced by a factor of 2
compared to the exact result. This is due to fluctuations of the
fermion charge $Q_i$\,.

\emph{Mean-field approximation:} The simplest approach to the
interacting system $J> 0$ is the Hartree approximation, i.e., magnetic
mean-field theory. This approximation does locally conserve the
auxiliary-fermion charge $Q_i$\,. Dyson's equation for the auxiliary
fermion reads
\begin{equation}
  \label{eqn-fdyson}
  \overline{G}_i(i\omega)
    = {}[\,i\omega + \mu^f - \overline{\Sigma}_i(i\omega)\,{}]^{-1}\;\;,
\end{equation}
and in Hartree approximation the self energy is independent of
$i\omega$ and given by
\begin{equation}
  \label{eqn-hartree}
  \overline{\Sigma}_i
    = \sum_j J_{ij}\frac{1}{2}\boldsymbol{\sigma}
      \underbrace{
        \textrm{Tr}^\sigma{}[\,
          \frac{1}{2}\boldsymbol{\sigma}
          \sum_{i\bar{\omega}}
          \overline{G}_j(i\bar{\omega})
          e^{i\bar{\omega} 0_+}\,{}]
          }_{\displaystyle = \langle \mathbf{S}_j\rangle}\;.
\end{equation}
Here the mean magnetization $\langle \mathbf{S}_j\rangle$ on the site
$j$ has been identified. For a square lattice with coordination number
$z= 4$ and only nearest-neighbor interaction $J>0$ we assume a
N{\'e}el state on the two sublattices $A$, $B$\,,
\begin{displaymath}
  \langle \mathbf{S}_A\rangle
    = - \langle \mathbf{S}_B\rangle
    = - \langle S^z_B\rangle\,\mathbf{e}_z\;.
\end{displaymath}
The fermion Green's function for any site on $A$ becomes
\begin{displaymath}
  \overline{G}_A(i\omega)
    = {}[\,i\omega + \mu^f + \sigma^z h\,{}]^{-1}
    \;\;,\;\;\;
  h= \frac{zJ}{2}\langle S^z_A\rangle\;,
\end{displaymath}
which leads to the self-consistent equation
\begin{equation}
  \label{eqn-selfhartree}
  \langle S^z_A\rangle
    = \frac{1}{2}{}[\,f(-h - \mu^f) - f(h - \mu^f)\,{}]\;.
\end{equation}
For average projection, with $\mu^f= 0$\,, we find
\begin{equation}
  \label{eqn-selfhartree-av}
  \textrm{average projection:} \quad\quad
  h= \frac{z}{4}J\,\tanh\big(\frac{h}{2T}\big)\;.
\end{equation}
Within the Popov--Fedotov scheme, using $\mu^f= i\frac{\pi}{2}T$\,,
one has
\begin{equation}
  \label{eqn-selfhartree-ex}
  \textrm{exact projection:} \quad\quad
  h= \frac{z}{4}J\,\tanh\big(\frac{h}{T}\big)\;,
\end{equation}
where the following expression for the Fermi function has been
utilized,
\begin{equation}
  \label{eqn-fermicmplx}
  f(x - i\frac{\pi}{2}T)
    = f(2x) + \frac{i}{2\cosh(x/T)}\;\;,
\end{equation}
for a real-valued $x$\,. In the physical observable
(\ref{eqn-selfhartree}) the imaginary part again cancels. Both
projection schemes lead to the same self-consistent equation for the
effective magnetic (Weiss) field $h$\,, except for a factor of 2 in
the temperature.  Accordingly the equations result in different
N{\'e}el temperatures,
\begin{displaymath}
  \begin{array}[c]{lcrcl}
    \textrm{average projection:} & & \displaystyle
      T_N & = & \displaystyle \frac{z}{8}J\;\;,
      \\[2ex]
    \textrm{exact projection:} & & \displaystyle
      T_N & = & \displaystyle \frac{z}{4}J\;.
  \end{array}
\end{displaymath}
However, at zero temperature both projection methods lead to the same
result,
\begin{equation}
  \label{eqn-hartreesatur}
  \textrm{average \emph{and} exact projection:} \quad
  \lim_{T\to 0}\langle S^z_A\rangle
    = \frac{1}{2}\;.
\end{equation}
That is, the unphysical reduction of the spin moment observed for the
free spin, is completely restored in the magnetically ordered ground
state. Apparently, the unphysical charge fluctuations are suppressed
at $T\to 0$\,.

\section{Effect of the constraint: self-consistent theory}
\label{sec-advanced}
The purpose of this section is to demonstrate the application of the
Popov--Fedotov scheme within a self-consistent approximation that goes
far beyond mean field. To our knowledge, the Popov--Fedotov approach
has at present been applied in mean-field-like calculations with
perturbative corrections\cite{popfed88,boukis99,kisopp00,dilric06},
but a self-consistent re-summation of the diagram series has not been
attempted.

When choosing an approximation scheme, it has to be kept in mind that
the automatic cancellation of unphysical states requires the
auxiliary-fermion charge $Q_i$ to be conserved (this has been
discussed in Section \ref{sec-intro} above). In particular, the local
gauge symmetry of the Hamitonian (\ref{eqn-allgrand}) under
$f_{i\alpha}\to e^{i \varphi_i}f_{i\alpha}$ must not be broken.
Accordingly, approximations leading to finite expectation values like
$\langle f^\dagger_{i\alpha} f_{j\alpha}\rangle\ne 0$ or $\langle
f_{i\uparrow} f_{j\downarrow}\rangle\ne 0$ cannot be used, while it is
safe to consider so-called $\Phi$-derivable
approximations\cite{lutwar60,bay62}.  Spontaneous breaking of physical
symmetries (e.g., spin rotation, lattice translation) may be included,
since the respective order parameters are gauge invariant. As a
consequence of the local gauge symmetry, the fermion Green's function
is always local,
\begin{equation}
  \label{eqn-local}
  \overline{G}_{ij}(i\omega)
    = \delta_{i,j}\,\overline{G}_{i}(i\omega)\;.
\end{equation}

Here we focus on the physics of the Heisenberg model
(\ref{eqn-hbergham}) in strictly two spatial dimensions at finite
temperature $T> 0$\,. This system has been studied extensively in the
past using a variety of numerical and analytical methods, in
particular in view of experiments on cuprate superconductors in the
undoped limit\cite{note-oldref}. The Hartree approximation discussed
in Section \ref{sec-simple} above is, of course, not appropriate for
the 2D system, although the N{\'e}el-ordered ground state at $T= 0$
appears to be qualitatively correct\cite{man91,briwol04}. At finite
$T$ the theorem of Mermin and Wagner requires the magnetization
$\langle \mathbf{S}_i\rangle= 0$ to vanish.  Therefore we seek an
approximation, where the susceptibility $\chi$ is self-consistently
coupled back onto itself. Such an approximation has originally been
proposed for the Hubbard model\cite{bicscawhi89}, and is commonly
referred to as FLEX\,.
\begin{figure}[tb]
  \centering
  \includegraphics[width=0.9\hsize]{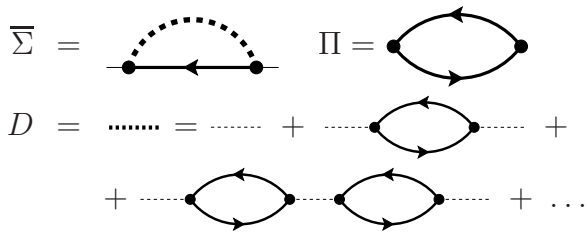}
  \caption{The self-consistent approximation (SCA) discussed in
    Section \ref{sec-advanced}\,, Eqs.(\ref{eqn-flex})\,. Continuous
    lines denote the fermion Green's function $\overline{G}$\,,
    Eq.(\ref{eqn-local})\,, the thin dashed line represents $J$\,, and
    dots are Pauli matrices $\times\,1/2$\,.}
  \label{fig-flex}
\end{figure}
For the Hamiltonian (\ref{eqn-allgrand}) with spin--spin interaction
it takes the form shown in Fig.\ \ref{fig-flex}\,. The fermion
self-energy shown in the figure reads
\begin{subequations}
\label{eqn-flex}
\begin{equation}
  \label{eqn-flex-self}
  \overline{\Sigma}_{i}(i\omega)
    = \frac{1}{4}
      T\sum_{i\nu}
       \sum_{\mu, \bar{\mu}}
       D_{ii}^{\mu\bar{\mu}}(i\nu)
       \,\sigma^\mu\,
       \overline{G}_i(i\omega + i\nu)
       \,\sigma^{\bar{\mu}}\;\;,
\end{equation}
$\omega$ and $\nu$ denote a fermionic and bosonic Matsubara frequency,
respectively, $\sigma^\mu$ a Pauli matrix, $\mu, \bar{\mu}= x, y,
z$\,. The renormalized spin--spin interaction on lattice sites $i, j$
is given by
\begin{equation}
  \label{eqn-flex-effint}
  D_{ij}^{\mu\bar{\mu}}(i\nu)
    = -J_{ij} \;+\;
      \sum_{l,k}
      J_{il}\,\chi_{lk}^{\mu\bar{\mu}}(i\nu)\,J_{kj}\;.
\end{equation}
The susceptibility $\chi$ represents the series of fermion-line
bubbles in Fig.\ \ref{fig-flex}\,,
\begin{equation}
  \label{eqn-flex-chi}
  \chi_{ij}^{\mu\bar{\mu}}
    = \big(\Pi\,{}[\,1 + J\Pi\,{}]^{-1}\big)_{ij}^{\mu\bar{\mu}}\;\;,
\end{equation}
with
\begin{equation}
  \label{eqn-flex-bubble}
  \Pi_i^{\mu\bar{\mu}}(i\nu)
    = -T\sum_{i\omega}
      \frac{1}{4}
      \textrm{Tr}^\sigma{}[\,\sigma^\mu\,
        \overline{G}_i(i\omega + i\nu)
        \,\sigma^{\bar{\mu}}\,
        \overline{G}_i(i\omega)\,{}]\;.
\end{equation}
\end{subequations}
In the paramagnetic phase with lattice-translational symmetry we have
\begin{displaymath}
  \overline{G}_i(i\omega)= \sigma^0\,G(i\omega)
    \;\;,\;\;\;
  \overline{\Sigma}_i(i\omega)= \sigma^0\,\Sigma(i\omega)\;\;,
\end{displaymath}
and therefore
\begin{displaymath}
  \Pi_i^{\mu\bar{\mu}}(i\nu)
    = \delta_{\mu, \bar{\mu}}\,\Pi(i\nu)\;.
\end{displaymath}
The self-consistent equations (\ref{eqn-flex}) with
Eq.(\ref{eqn-fdyson}) now turn into
\begin{subequations}
\label{eqn-sflex}
\begin{eqnarray}
  \label{eqn-sflex-bubble}
  \Pi(i\nu)
    & = & -\frac{T}{2}\sum_{i\omega}
          G(i\omega + i\nu)\,G(i\omega)\;\;,
      \\
  \label{eqn-sflex-chi}
  \chi(\mathbf{q}, i\nu)
    & = & \frac{\Pi(i\nu)}{1 + J(\mathbf{q}) \Pi(i\nu)}\;\;,
      \\
  \label{eqn-sflex-effint}
  D(i\nu)
    & = & \frac{1}{N_L}
      \sum_{\mathbf{q}}
      J^2(\mathbf{q})\,\chi(\mathbf{q}, i\nu)\;\;,
      \\
  \label{eqn-sflex-self}
  \Sigma(i\omega)
    & = & \frac{3T}{4}\sum_{i\nu}
      D(i\nu)\,G(i\omega + i\nu)\;\;,
      \\
  \label{eqn-sflex-dyson}
  G(i\omega)
    & = & \big[\,i\omega + \mu^f - \Sigma(i\omega)\,\big]^{-1}\;.
\end{eqnarray}
The bare interaction in wave-vector space reads, for a square lattice
in 2D with nearest-neighbor distance $a\equiv 1$\,,
\begin{equation}
  \label{eqn-sflex-jj}
  J(\mathbf{q})
    = 4J\,\gamma(\mathbf{q})  \;\;,\;\;\;
  \gamma(\mathbf{q})
    = \frac{1}{2}{}[\,\cos(q_x) + \cos(q_y)\,{}]\;.
\end{equation}
\end{subequations}
Note that the bare $J$ in Eq.(\ref{eqn-flex-effint}) does not
contribute to the local $D(i\nu)\equiv D_{ii}(i\nu)$\,,
Eq.(\ref{eqn-sflex-effint})\,, since $J_{ii}= 0$\,.

It has been emphasized above, that the fermion propagator and as a
consequence the irreducible bubble are local, $G_{ij}=
\delta_{i,j}\,G$\,, $\Pi_{ij}= \delta_{i,j}\,\Pi$\,. Nevertheless, the
interesting measurable\cite{foot-cons} quantity in the
Eqs.(\ref{eqn-sflex}) is the susceptibility $\chi(\mathbf{q})$\,,
which is wave-vector dependent through the bare interaction
$J(\mathbf{q})$\,, and therefore may describe even long-range
fluctuations.

The SCA shown in Fig.\ \ref{fig-flex} can be derived from a
$\Phi$-functional in close analogy\cite{bicscawhi89} to the FLEX\,.
Therefore, Eqs.(\ref{eqn-sflex}) represent a conserving approximation
and can be used with the averaged fermion constraint as well as the
Popov--Fedotov approach. The case of average projection, $\mu^f= 0$\,,
has been treated in detail in Ref.\ \onlinecite{briwol04}\,. The
magnetic correlation length $\xi(T)$ and the dynamical structure
factor, derived from the susceptibility (\ref{eqn-sflex-chi})\,, came
out quite satisfactorily when compared to known results, indicating
that the diagrams in Fig.\ \ref{fig-flex} indeed capture the important
physics of the 2D system at low $T$\,. In the following some of the
results from Ref.\ \onlinecite{briwol04} will be re-calculated using
the Popov--Fedotov method, i.e., $\mu^f= i\frac{\pi}{2}T$\,. It will
turn out, that the imaginary-valued chemical potential requires the
use of complex-valued spectral functions, leading to more involved
equations than those derived in Ref.\ \onlinecite{briwol04} for
$\mu^f= 0$\,.  The results calculated with both methods differ, except
in the limit of vanishing temperature, where average and exact
projection become equal, as will be presented below.

\emph{Equations for exact projection:} Before discussing the numerical
solution of Eqs.(\ref{eqn-sflex}), we quote some important formal
results derived in Appendix \ref{sec-app-greens}\,. For a numerical
implementation it is suitable to express all Green's functions through
their respective spectral functions. On account of the Hamitonian
being non-Hermitian, the spectral function of the fermion Green's
function $G(i\omega)$ becomes complex-valued. $G$ is given by
\begin{equation}
  \label{eqn-gfdef}
  (\overline{G}_i)^{\alpha\bar{\alpha}}(i\omega)
    = \int_0^{1/T}\mathrm{d}\tau\,
      e^{i\omega\tau}
      \langle\mathcal{T}_\tau\{f_{i\alpha}(\tau)
        f^\dagger_{i\bar{\alpha}}(0)\}\rangle^\ff\;\;,
\end{equation}
with the thermal expectation value and $\tau$-dependence calculated in
the enlarged fermion Fock-space with the Hamiltonian
(\ref{eqn-allgrand})\,. $G$ has the following spectral representation,
$\displaystyle
  \overline{G}^{\alpha\bar{\alpha}}_i(i\omega)
    = \delta_{\alpha, \bar{\alpha}}
      G(i\omega)$\,,
\begin{equation}
  \label{eqn-gfspect}
  G(i\omega)
    = \int_{-\infty}^\infty\mathrm{d}\varepsilon\,
      \frac{\widehat{G}(\varepsilon)}{i\omega + \mu^f - \varepsilon}
      \;\;,\;\;\;
  \mu^f = i\frac{\pi}{2}T\;.
\end{equation}
The energy variable $\varepsilon$ is a real number. For simplicity a
system invariant under lattice translations and spin rotations has
been assumed. The spectral function $\widehat{G}$ is complex-valued,
\begin{equation}
  \label{eqn-fspect}
  \widehat{G}(\varepsilon)
    = \rho_1(\varepsilon) \;+\;
      i\rho_2(\varepsilon)\;.
\end{equation}
In the Appendix \ref{sec-app-greens} we also derive the sum rule
\begin{equation}
  \label{eqn-fsum}
  \int\mathrm{d}\varepsilon\,
    \widehat{G}(\varepsilon)= 1\;\;,
\end{equation}
\begin{displaymath}
  \;\;\;\Rightarrow\;\;\; 
  \int\mathrm{d}\varepsilon\,
    \rho_1(\varepsilon)= 1  \;\;,\;\;\;
  \int\mathrm{d}\varepsilon\,
    \rho_2(\varepsilon)= 0\;\;,
\end{displaymath}
$\varepsilon$ is again real valued. For the special case of spin
degeneracy considered here, the fermion spectral-function in addition
obeys the ``particle--hole'' symmetry,
\begin{equation}
  \label{eqn-phsymm}
  \widehat{G}(-\varepsilon)
    = \widehat{G}(\varepsilon)^\ast\;\;,
\end{equation}
\begin{displaymath}
  \;\;\;\Rightarrow\;\;\;
  \rho_1(-\varepsilon)= \rho_1(\varepsilon)  \;\;,\;\;\;
  \rho_2(-\varepsilon)= - \rho_2(\varepsilon)\;.
\end{displaymath}
The relations (\ref{eqn-gfspect}), (\ref{eqn-fspect}),
(\ref{eqn-phsymm}) hold as well for the fermion self-energy
$\Sigma$\,.

For a numerical solution of the Eqs.(\ref{eqn-sflex}), we introduce
structure factors for $\Pi(i\nu)$ and $D(i\nu)$ according to
\begin{displaymath}
  S^0(\omega)
    = {}[\,1 + g(\omega)\,{}]\,\Pi''(\omega)
      \;\;,\;\;\;
  \Pi''(\omega)
    = \textrm{Im}\,\Pi(\omega + i0_+)\;\;,
\end{displaymath}
\begin{displaymath}
  U(\omega)
    = {}[\,1 + g(\omega)\,{}]\,
      \textrm{Im}\,D(\omega + i0_+)\;\;,
\end{displaymath}
with analytic continuation to the real axis via
$\displaystyle
  \Pi(i\nu)\to \Pi(\omega + i0_+)$
and
$\displaystyle
  D(i\nu)\to D(\omega + i0_+)$\,,
$\displaystyle
  \omega\in\mathbb{R}$\,.
  The Bose function is denoted by $g(\omega)= 1/(e^{\omega/T} - 1)$\,.
  As shown in detail in Appendix \ref{sec-app-deriv}\,,
  Eqs.(\ref{eqn-sflex}) now turn into
\begin{subequations}
  \label{eqn-nflex}
\begin{eqnarray}
  U(\omega)
    & = &  \label{eqn-nflex-effint}
      \int_{-\infty}^\infty\mathrm{d}\varepsilon\,
      \frac{S^0(\omega)\,\,\mathcal{N}(\varepsilon)\,\varepsilon^2}
           {{}\big(1 + \varepsilon\,\Pi'(\omega){}\big)^2 +
            {}\big(\varepsilon\,\Pi''(\omega){}\big)^2}\;\;,
      \\
  \Pi''(\omega)
    & = &  \label{eqn-nflex-piim}
      S^0(\omega) - S^0(-\omega)\;\;,
      \\
  \Pi'(\omega)
    & = &  \label{eqn-nflex-pire}
      \textrm{P}\!\!\int_{-\infty}^\infty
      \frac{\mathrm{d}\varepsilon}{\pi}
      \frac{\Pi''(\varepsilon)}{\varepsilon - \omega}\;\;,
      \\
  S^0(\omega)
    & = &  \label{eqn-nflex-snul}
      \frac{\pi}{2}
      \int_{-\infty}^\infty\mathrm{d}\varepsilon\,
        \widehat{G}^+(\varepsilon)\,
        \widehat{G}^-(\varepsilon - \omega)\;\;,
\end{eqnarray}
for the real (physical) functions $U, S^0, \Pi'', \Pi'$ with the
Kramers--Kroenig transform $\Pi'$\,, and
\begin{eqnarray}
  \widehat{G}^+(\omega)
    & = &  \label{eqn-nflex-gfplus}
      {}[\,1 - f(\omega - i\frac{\pi}{2}T)\,{}]
      \,\widehat{G}(\omega)\;\;,
      \\
  \widehat{G}^-(\omega)
    & = &  \label{eqn-nflex-gfmin}
      f(\omega - i\frac{\pi}{2}T)
      \,\widehat{G}(\omega)\;\;,
      \\
  \widehat{G}(\omega)
    & = &  \label{eqn-nflex-gf}
      \frac{\widehat{\Sigma}(\omega)}
           {{}\big(\omega - \overline{\Sigma}(\omega)\big)^2 +
            {}\big(\pi\widehat{\Sigma}(\omega)\big)^2}\;\;,
      \\
  \widehat{\Sigma}(\omega)
    & = &  \label{eqn-nflex-selfspec}
      \frac{3}{4\pi}
      \int_{-\infty}^\infty\mathrm{d}\varepsilon\,
      U(\varepsilon)\,
      \big[\,\widehat{G}^+(\omega - \varepsilon) +
      \\
    &   &  \nonumber\hspace*{2cm}\mbox{}+
        \widehat{G}^-(\omega + \varepsilon)\,\big]\;\;,
      \\
  \overline{\Sigma}(\omega)
    & = &  \label{eqn-nflex-selfhil}
      \textrm{P}\!\!\int_{-\infty}^\infty\mathrm{d}\varepsilon\,
      \frac{\widehat{\Sigma}(\varepsilon)}
           {\omega - \varepsilon}\;\;,
\end{eqnarray}
\end{subequations}
for the complex (unphysical) spectra $\widehat{G}^\pm, \widehat{G},
\widehat{\Sigma}, \overline{\Sigma}$ with the Hilbert transform
$\overline{\Sigma}$\,. Note again that the energy arguments $\omega,
\varepsilon$ are real-valued. As is also shown in the Appendix
\ref{sec-app-deriv}\,, the Eqs.(\ref{eqn-nflex}) can be somewhat
simplified by utilizing the symmetry (\ref{eqn-phsymm})\,. The
imaginary chemical potential $i\frac{\pi}{2}T$ appears in
Eq.(\ref{eqn-nflex-gfplus}) and (\ref{eqn-nflex-gfmin}), adding an
imaginary part to the fermion spectra $\widehat{G}^+$ and
$\widehat{G}^-$ via Eq.(\ref{eqn-fermicmplx})\,.

The density-of-states that enters Eq.(\ref{eqn-nflex-effint}) is
defined as
\begin{equation}
  \label{eqn-dos}
  \mathcal{N}(\varepsilon)
    = \frac{1}{N_L}\sum_{\mathbf{k}}
      \delta(\varepsilon - J(\mathbf{q}))\;\;,
\end{equation}
and for the nearest-neighbor interaction (\ref{eqn-sflex-jj}) in two
dimensions it becomes
\begin{equation}
  \label{eqn-dosnum}
  \mathcal{N}(\varepsilon)
    = \frac{K(m)}{2\pi^2 J}
      \theta(4J - |\varepsilon|)
      \;\;,\;\;\;
  m = 1 - \left(\frac{\varepsilon}{4J}\right)^2\;\;,
\end{equation}
with the complete elliptic integral of the first kind, $K(m)=
\int_0^1\mathrm{d}t\,{}[\,(1 - t^2)(1 - m t^2)\,{}]^{-1/2}$\,.

The physical output from the numerical iteration of
Eqs.(\ref{eqn-nflex}), (\ref{eqn-dosnum}) is the structure factor
$U(\omega)$ of the effective interaction, which is essentially the
local (on-site) spin-excitation spectrum, and the wave-vector
dependent dynamical spin-structure factor
\begin{equation}
  \label{eqn-struct}
  S(\mathbf{q}, \omega)
    = {}[\,1 + g(\omega)\,{}]\,
      \textrm{Im}
      \left\{\frac{\Pi(\omega)}{1 + J(\mathbf{q})\Pi(\omega)}
        \right\}\;\;,
\end{equation}
with
$\displaystyle
  \Pi= \Pi' + i\Pi''$\,.
Furthermore, the magnetic correlation length $\xi(T)$ is extracted
from the static magnetic susceptibility $\chi(\mathbf{q}, 0)$\,:
Eq.(\ref{eqn-sflex-chi}), for $\mathbf{q}$ close to the
N{\'e}el-ordering vector $\mathbf{Q}= (\pi,\pi)$\,, i.e., 
\begin{displaymath}
  J(\mathbf{q})
    \simeq -4J + (\mathbf{q} - \mathbf{Q})^2J\;\;,
\end{displaymath}
leads to
\begin{displaymath}
  \chi(\mathbf{q}, 0)
    = \frac{\Pi'(0)}{1 + J(\mathbf{q})\Pi'(0)}
    \simeq \frac{1}{J}
      \frac{1}{\xi^{-2} + (\mathbf{q} - \mathbf{Q})^2}\;\;,
\end{displaymath}
where $\Pi(0)= \Pi'(0)$ has been used, and the correlation length is
identified as
\begin{equation}
  \label{eqn-xidef}
  \xi(T)
    = \left(\frac{J\Pi'(0)}{1 - 4J\Pi'(0)}\right)^{1/2}\;.
\end{equation}
For completeness, in Appendix \ref{sec-app-simpl} the self-consistent
equations (\ref{eqn-nflex}) are re-written in real-valued spectral
functions. These Eqs.(\ref{eqn-simpl-num}) can directly be compared to
the Eqs.(A1) in Ref.\ \onlinecite{briwol04}\,: Both sets of equations
represent the same diagrammatic approximation shown in Fig.\
\ref{fig-flex}\,, the former derived within the Popov--Fedotov scheme
(exact projection), the latter within average projection.

\section{Numerical results}
\label{sec-num}
The numerical results presented below are obtained from an iterative
solution of the self-consistent equations (\ref{eqn-sflex}), using the
identical procedure and parameters for exact projection ($\mu^f=
i\frac{\pi}{2}T$\,, leading to Eqs.(\ref{eqn-simpl-num}) in App.\
\ref{sec-app-simpl}) as well as average projection ($\mu^f= 0$\,,
leading to Eqs.(A1) in Ref.\ \onlinecite{briwol04}). The procedure
also utilizes Eqs.(\ref{eqn-dosnum}) and (\ref{eqn-xidef}), which
apply to both projection schemes. The data shown in Ref.\
\onlinecite{briwol04} for the case of average projection have not been
re-used in the present study.

\begin{figure}[tb]
  \centering
  \includegraphics[width=\hsize,clip=true]{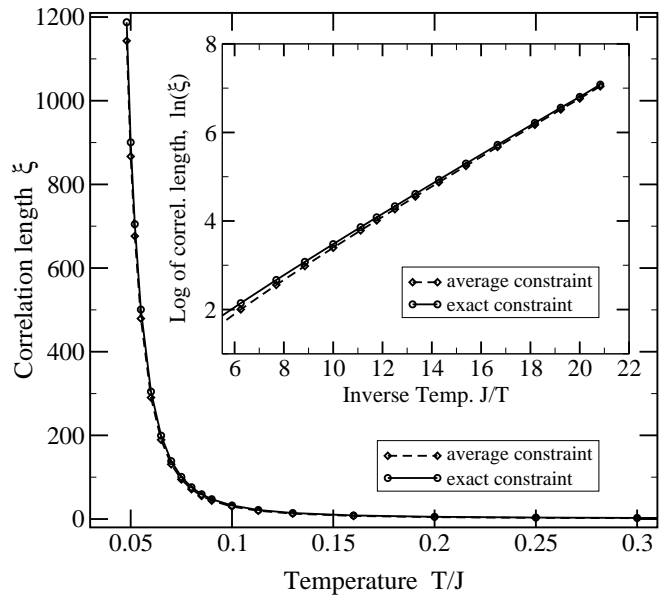}
  \caption{\textbf{Main figure:} Magnetic correlation length $\xi(T)$
    from the numerical solution of the self-consistent approximation
    Fig.(\ref{fig-flex}) for the 2D square lattice, in units of the
    lattice spacing. The continuous line with diamonds corresponds to
    Eqs.(\ref{eqn-nflex}), where the Popov--Fedotov approach has been
    used. The dashed line with circles belongs to Eqs.(A1) in Ref.\
    \onlinecite{briwol04}\,, where the auxiliary-particle constraint
    has been approximated by its thermal average. \textbf{Inset:} The
    correlation length, plotted as $\ln(\xi)$ vs.\ $J/T$\,. Shown is
    the temperature region $0.048\,J\le T\le 0.16\,J$\,, where a fit
    to the data has been performed as described in the text.}
  \label{fig-correl}
\end{figure}

\emph{Magnetic correlation length:} The correlation length $\xi(T)$ is
shown in Fig.\ \ref{fig-correl}\,.  $\xi(T)$ becomes larger than one
lattice spacing for $T\lesssim J$ and reaches values up to $\simeq
1200$ lattice spacings for the lowest temperature $T= 0.048 J$
considered in this work. The numerical data shown in Fig.\
\ref{fig-correl} are well reproduced by
\begin{equation}
  \label{eqn-fit}
  \xi(T)
    = c\left(\frac{J}{T}\right)^b
      \exp\big(a\,\frac{J}{T}\big)\;.
\end{equation}
The parameters $a, b, c$ are determined by plotting the data as
$\ln(\xi)$ vs.\ $J/T$ (see the inset of Fig.\ \ref{fig-correl}) and a
numerical fit of the function $\;\ln(c) + b\,\ln(J/T) + a\,J/T\;$ to
the data. We find
\begin{equation}
  \label{eqn-fitval}
  \begin{array}[c]{lrcl}
    \textrm{average projection:}
      & a & = & 0.296  \\
      & b & = & 0.592  \\
      & c & = & 0.393  \\ \\
    \textrm{exact projection:}
      & a & = & 0.304  \\
      & b & = & 0.418  \\
      & c & = & 0.595
  \end{array}
\end{equation}
At low temperatures the effect of the exact auxiliary-particle
projection is only marginal, as is already apparent from inspecting
Fig.\ \ref{fig-correl}\,. The ``spin stiffness'' $a= 2\pi\rho_s/J$ in
the exponent in Eq.(\ref{eqn-fit}) is the same in both
cases\cite{note-num}, $a\simeq 0.30$\,, merely the power $b\simeq 0.5$
of the algebraic prefactor is slightly modified in going from average
to exact projection.

\begin{figure}[tb]
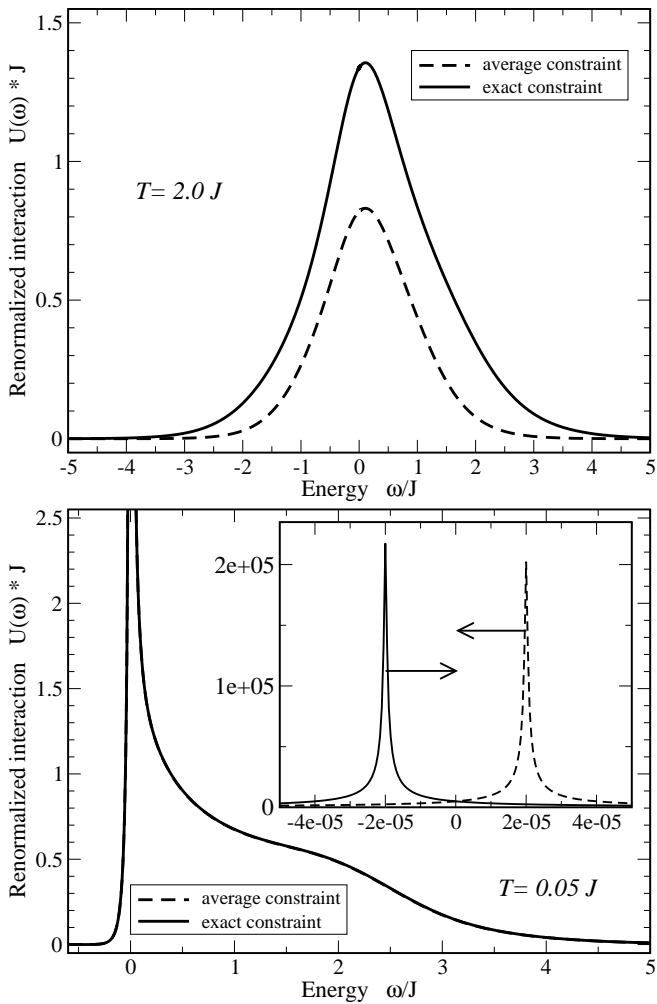

  \centering
  \includegraphics[width=\hsize,clip=true]{figure3top.eps}
  \includegraphics[width=\hsize,clip=true]{figure3bottom.eps}
  \caption{\textbf{Top:} The structure factor $U(\omega)$\,,
    Eq.(\ref{eqn-nflex-effint}), of the renormalized local interaction
    $D(i\nu)$\,, Eq.(\ref{eqn-sflex-effint}), at high temperature $T=
    2J$\,. Dashed and continuous lines as in Fig.\ \ref{fig-correl}\,.
    \textbf{Bottom, main figure:} $U(\omega)$ at low temperature $T=
    0.05J$\,, corresponding to a correlation length $\xi\simeq 900$\,.
    \textbf{Bottom, inset:} Same data as in the main figure, but
    zoomed around the energy $\omega= 0$\,. For clarity, the peaks for
    average constraint (dashed) and exact constraint (continuous) are
    shifted from their original position $\omega= 0$ by
    \mbox{2e-05}$J$ and \mbox{-2e-05}$J$\,, respectively.  }
  \label{fig-effint}
\end{figure}

\emph{Spin spectral-function and energy scale:} The almost vanishing
effect of the auxiliary-particle constraint at low temperature is also
visible in the spectra: Fig.\ \ref{fig-effint} shows the effective
interaction $U(\omega)$\,, Eq.(\ref{eqn-nflex-effint}), which is the
structure factor of the $D(i\nu)$ given in
Eq.(\ref{eqn-sflex-effint})\,. Since $J(\mathbf{q})^2$ in
Eq.(\ref{eqn-sflex-effint}) depends only weakly on $\mathbf{q}$\,,
$U(\omega)$ is essentally the local magnetic structure factor or
spin-excitation spectrum. Note that Eq.(\ref{eqn-nflex-effint}) is the
same in the average-projected case, Eq.(A1h) in Ref.\
\onlinecite{briwol04}\,. The data for low temperature, shown in the
bottom panel of Fig.\ \ref{fig-effint}\,, features a broad shoulder of
width $\sim J$\,, which is reminiscent of the box-like
density-of-states for spin waves in 2D\,. Around zero energy
$U(\omega)$ displays a huge peak (see the inset of the figure), which
contains the critical fluctuations at $\mathbf{q}\simeq (\pi,\pi)$
close to the antiferromagnetic ordering wave-vector. The curves for
average and exactly treated constraint are on top of each other;
merely the amplitudes of the peaks at $\omega= 0$ differ by a factor
of $\mathcal{O}(1)$\,. At high temperature $T= 2J$\,, on the contrary,
the curves for average and exact constraint are well separated, in
particular the total spectral weight is smaller for average
projection. This can be seen from the data in the top panel of Fig.\
\ref{fig-effint}\,. The reduction of spectral weight in $U(\omega)$ is
related to an unphysical lack of local spin moment, which occurs if
the constraint is not taken exactly. This will be discussed further
below.
\begin{figure}[tb]
  \centering
  \includegraphics[width=\hsize,clip=true]{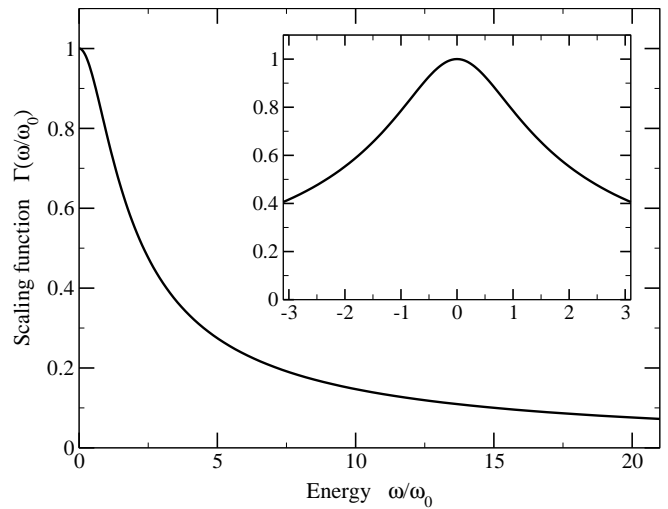}
  \caption{\textbf{Main figure:} The scaling function
    Eq.(\ref{eqn-scalefun}) of the renormalized spin--spin
    interaction, using the energy scale $\omega_0(T)$ from
    Eq.(\ref{eqn-enscale}) and $\Gamma(0)\equiv 1$\,. The curve shown
    here is the same for all temperatures $0.048\le T/J\le 0.13$\,,
    corresponding to $1000\ge \xi(T)\ge 10$\,, and independent of the
    auxiliary-particle constraint being treated exactly or on average.
    \textbf{Inset:} Detailed view of the region near $\omega= 0$\,.}
  \label{fig-enscale}
\end{figure}
\begin{figure}[tb]
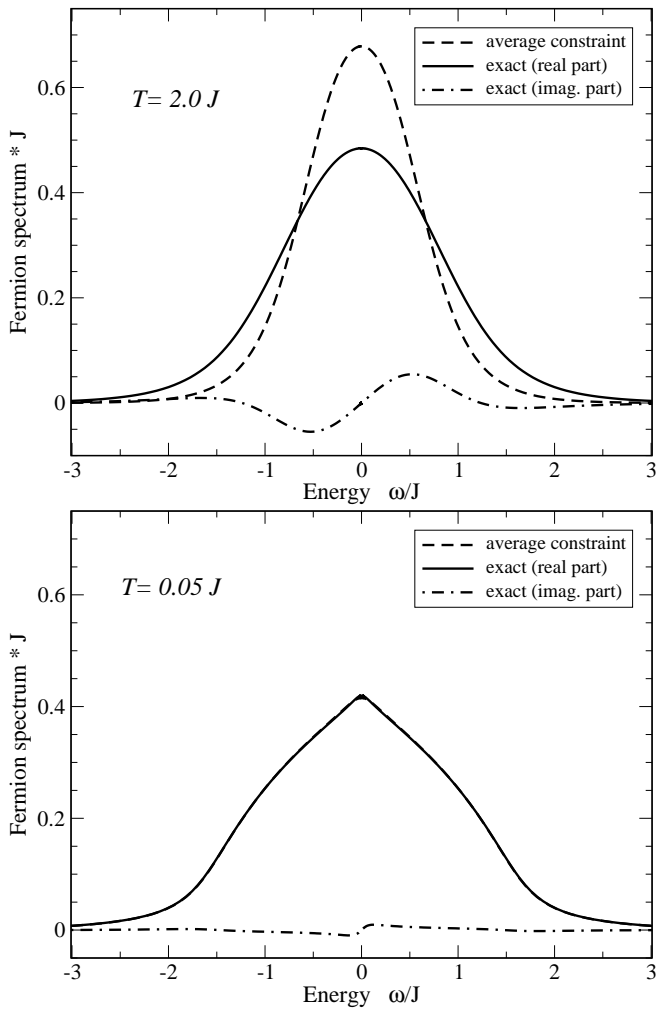

  \centering
  \includegraphics[width=\hsize,clip=true]{figure5top.eps}
  \includegraphics[width=\hsize,clip=true]{figure5bottom.eps}
  \caption{The spectral function $\widehat{G}(\omega)$ of the
    auxiliary fermion, introduced in Eqs.(\ref{eqn-gfdef}),
    (\ref{eqn-gfspect})\,.  For exact auxilary-fermion projection it
    is complex valued and given by Eq.(\ref{eqn-nflex-gf}), for
    average projection Eq.(A1e) from Ref.\ \onlinecite{briwol04}
    holds, with $\widehat{G}(\omega)$ being real. \textbf{Top:} high
    temperature, \textbf{Bottom:} low temperature. Parameters as in
    Fig.\ \ref{fig-effint}\,.}
  \label{fig-fdos}
\end{figure}

The difference of peak amplitudes visible in the inset of Fig.\
\ref{fig-effint} bottom can be traced back to the slightly different
correlation length $\xi(T)$\,, compare Eq.(\ref{eqn-fitval})\,. The
influence of the absolute value of $\xi$ vanishes if the scaling
behaviour of $U(\omega)$ is considered: We start from the dynamical
scaling hypothesis\cite{chn89},
\begin{equation}
  \label{eqn-hypo}
  S(\mathbf{q}, \omega)
    = \frac{1}{\omega_0}
      S^{st}(\mathbf{Q})
      \,\varphi(k\xi)\,\Phi(k\xi, \omega/\omega_0)\;.
\end{equation}
Here $S(\mathbf{q}, \omega)$ is the dynamical structure factor
Eq.(\ref{eqn-struct}), $S^{st}(\textbf{Q})$ denotes the static
(equal-time) correlation function
\begin{equation}
  \label{eqn-statstruct}
  S^{st}(\mathbf{q})
    = \langle S^x_{\mathbf{q}} S^x_{-\mathbf{q}}\rangle
    = \frac{1}{\pi}
      \int_{-\infty}^\infty\mathrm{d}\omega\,
      S(\mathbf{q}, \omega)
      \;\;,\;\;\; 
\end{equation}
taken at the AF ordering wave vector $\mathbf{Q}= (\pi,\pi)$\,.
$\varphi(x)$\,, $\Phi(x, y)$ are scaling functions, $\mathbf{k}=
\mathbf{q} - \mathbf{Q}$\,, and $\omega_0$ is the energy scale for
critical fluctuations. At small energies $\omega\ll J$ we expect
$U(\omega)\propto\int\mathrm{d}^2q\,S(\mathbf{q}, \omega)$\,, and from
an intergration of Eq.(\ref{eqn-hypo}) over wave-vector space there
follows the scaling property
\begin{equation}
  \label{eqn-lochypo}
  U(\omega)
    = \frac{1}{\omega_0}
      S^{st}(\mathbf{Q})
      \,\Gamma(\omega/\omega_0)\;.
\end{equation}
$\Gamma(y)$ is the (a priori unknown) scaling function for the local
effective spin-spin interaction $U(\omega)$\,. According to
Eq.(\ref{eqn-hypo}) the energy scale can be extracted from the
numerical data, up to a constant prefactor, using
\begin{equation}
  \label{eqn-enscale}
  \omega_0
    = \left.\frac{S^{st}(\mathbf{Q})}{S(\mathbf{Q}, \omega)}
      \right|_{\omega= 0}\;.
\end{equation}
We obtained the energy scale in the temperature range $0.048\le T/J\le
0.13$\,, which corresponds to correlation lengths $1000\ge \xi\ge
10$\,, using Eqs.(\ref{eqn-enscale}), (\ref{eqn-statstruct}), and
(\ref{eqn-struct})\,. The scaling function is then determined for each
temperature from Eq.(\ref{eqn-lochypo}),
\begin{equation}
  \label{eqn-scalefun}
  \Gamma(\omega/\omega_0)
    = \frac{U(\omega)}{U(0)}\;.
\end{equation}
All curves $\Gamma(\omega/\omega_0)$\,, whether calculated with
average or with exact auxiliary-particle constraint, agree to within
numerical accuracy. The scaling function is shown in Fig.\
\ref{fig-enscale}\,. Deviations from scaling behaviour become visible
only for higher energies $\omega/\omega_0> 20$\,. In particular, going
from exact to average projection has no effect on the scaling
behavior.

A slight difference in the low-temperature properties of the two
auxiliary-particle methods shows up in the energy scale itself: From a
linear regression on $\omega_0(T)\times\xi^2(T)$\,, obtained from
Eq.(\ref{eqn-enscale}) and Eq.(\ref{eqn-xidef}), we find
\begin{equation}
  \label{eqn-numenscale}
  \begin{array}[c]{lrcl}
    \textrm{average:}
      & \omega_0 \xi^2 & = & (0.234 \pm 0.002)\,J\;\;,
        \\
    \textrm{exact:}
      & \omega_0 \xi^2 & = & (0.234 \pm 0.002)\,J \;+\; 0.089\,T\;.
  \end{array}
\end{equation}
Nevertheless, the temperature behavior of the energy scale essentially
is the same,
\begin{equation}
  \label{eqn-enscaleresult}
  \omega_0(T)
    = 0.23\cdot J \,(\xi)^{-2}  \;\;,\;\;\; 
\end{equation}
which corresponds to a dynamical critical exponent $z= 2$\,.

\emph{Fermion propagator:} Fig.\ \ref{fig-fdos} display the spectral
function of the auxiliary-fermion propagator for the two projection
methods. As has been discussed below Eq.(\ref{eqn-gfdef}), the
spectrum $\widehat{G}(\omega)$ of the fermion is complex valued, if
the auxiliary-particle constraint is enforced exactly via the
imaginary-valued chemical potential $\mu^f= i\frac{\pi}{2}T$\,. If the
constraint is treated on the average using $\mu^f= 0$\,,
$\widehat{G}(\omega)$ remains real. In Fig.\ \ref{fig-fdos} the
spectrum $\widehat{G}(\omega)$ is shown for high (top panel) and low
temperature (bottom). At high $T$ the spectra for exact and average
constraint differ significantly, in particular the
$\widehat{G}(\omega)$ from exact projection has a considerable
imaginary part. At low temperature, on the other hand, the imaginary
part is quite small, while the real part becomes equal to the spectrum
of the average projection. In accordance with the results for
spin-structure factor and correlation length, the projection onto the
physical part of the fermion-Hilbert space has almost no effect at
sufficiently low temperature.

\begin{figure}[tb]
  \centering
  \includegraphics[width=\hsize,clip=true]{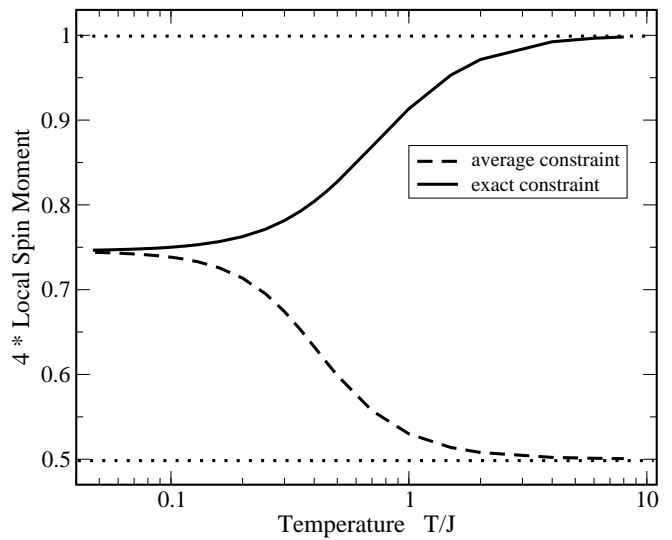}
  \caption{The local spin moment $S^{st}_{loc}$ as defined in
    Eq.(\ref{eqn-momdef}), multiplied by 4\,. The limiting values for
    $T\gg J$ given in Eq.(\ref{eqn-momhigh}) are indicated as dotted
    lines. The exact value, given by the sum rule (\ref{eqn-momrule}),
    is $4\,S^{st}_{loc}= 1$\,. }
  \label{fig-locmom}
\end{figure}

\emph{Local spin moment:} Another interesting quantity for studying
the influence of the auxiliary-fermion constraint is the local spin
moment, given by the local equal-time correlation function at an
arbitrary site $i$\,,
\begin{equation}
  \label{eqn-momdef}
  S^{st}_{loc}
    = \langle S^x_i S^x_i\rangle
    = \frac{1}{N_L}\sum_{\mathbf{q}}S(\mathbf{q})\;.
\end{equation}
The static structure factor $S(\mathbf{q})$ has been defined in
Eq.(\ref{eqn-statstruct})\,. For a spin-1/2 system in the paramagnetic
phase $S^{st}_{loc}$ should fulfill the sum rule
\begin{equation}
  \label{eqn-momrule}
  S^{st}_{loc}
    = \frac{1}{3}(\mathbf{S}_i)^2
    = \frac{1}{3}S(S + 1)
    = \frac{1}{4}\;.
\end{equation}
At very high temperature $T\gg J$\,, interaction effects can be
ignored, and $S^{st}_{loc}$ is given by the simple bubble $\Pi$ shown
in Fig.\ \ref{fig-flex}\,, calculated with free auxiliary fermions.
With Eq.(\ref{eqn-sflex-bubble}) we have
\begin{eqnarray*}
  T\gg J\,: \quad
  S^{st}_{loc}
    & \simeq & T\sum_{i\nu}\Pi(i\nu)
      \big|_{J= 0}
      \\
    & = & \frac{1}{2}
      \langle f^\dagger_{i\uparrow} f_{i\uparrow}\rangle^\ff\,
      \langle f_{i\downarrow} f^\dagger_{i\downarrow}\rangle^\ff
      \big|_{J= 0}\;.
\end{eqnarray*}
For both projection methods, the expectation values are to be taken in
the generalized grand-canonical ensemble (\ref{eqn-allgrand}),
(\ref{eqn-chem})\,. For the case of average projection we find
$\langle f^\dagger_{i \sigma} f_{i \sigma}\rangle= f(0)= 1/2$ for any
spin direction $\sigma= \pm 1$\,, with exact projection we have
$\langle f^\dagger_{i \sigma} f_{i \sigma}\rangle=
f(-i\frac{\pi}{2}T)= \frac{1}{2}(1 + i)$\,.  $f$ denotes the Fermi
function. That is, for $T\gg J$\,,
\begin{equation}
  \label{eqn-momhigh}
  \begin{array}[c]{lrcl}
    \textrm{average constraint:} &
      4\,S^{st}_{loc} & = & 
        \frac{1}{2}\;\;,
        \\
    \textrm{exact constraint:} &
      4\,S^{st}_{loc} & = & 
        2\,\left|\frac{1}{2}(1 + i)\right|^2= 1\;.
  \end{array}
\end{equation}
If the projection onto the physical Hilbert space is performed
exactly, the sum rule (\ref{eqn-momrule}) is correctly reproduced.
With average projection, on the other hand, it is significantly
violated. This is due to thermal charge fluctuations into unphysical,
spinless states, which reduce the spin moment.

When temperature is lowered, we find a partly unexpected result: Fig.\
\ref{fig-locmom} shows $4\,S^{st}_{loc}$ from the numerical solution
as function of $T$\,. At high temperature the free-spin result is
approached, whereas at low temperature the average and exact
projection methods lead to the same value for $S^{st}_{loc}$\,. This
observation fits into the line of results obtained so far: at $T\to 0$
it does not matter whether the auxiliary-particle constraint is
treated exactly or on the thermal average.

However, the local spin moment at $T\to 0$\,, $4\,S^{st}_{loc}\simeq
0.75$ is too small. This is not due to an ill-treated constraint, but
an artifact of the approximation to the interacting system. The local
moment (\ref{eqn-momdef}) measures the total spectral weight of spin
excitations, averaged over the Brillouin zone. The self-consistent
approximation we use, see Fig.\ \ref{fig-flex}\,, apparently lacks
some weight in the spin-excitation spectrum. In Ref.\
\onlinecite{briwol04} we studied an approximation with a somewhat
reduced self consistency (called ``MSCA''), which delivered a better
result for $S^{st}_{loc}$ at low $T$\,, $4\,S^{st}_{loc}\simeq
0.85$\,.  Moreover, the wave-vector dependence of $S(\mathbf{q},
\omega)$ came out better. However, the MSCA cannot straight-forwardly
be extended to the Popov--Fedotov scheme, since we don't know a
$\Phi$-functional for that approximation to guarantee the conservation
of the auxiliary-charge $Q_i$\,, which is a necessary condition for
the Popov--Fedotov method (see Section \ref{sec-intro} above).

\emph{The limit $T\to 0$\,:} At first sight it seems trivial that the
imaginary-valued chemical potential $\mu^f= i\frac{\pi}{2}T$ has
almost no effect at $T\to 0$\,: $\mu^f$ enters the self-consistent
equations through the Fermi function $f(\omega - i\frac{\pi}{2}T)$ in
Eqs.(\ref{eqn-nflex-gfplus}) and (\ref{eqn-nflex-gfmin})\,. Assuming
$T\ll\omega$\,, Eq.(\ref{eqn-fermicmplx}) yields
\begin{subequations}
  \label{eqn-flimit}
\begin{equation}
  \label{eqn-flimit-low}
  T\ll\omega\,: \quad
  f(\omega - i\frac{\pi}{2}T)
    = \Theta(-\omega) + \mathcal{O}(e^{-2|\omega|/T})\;\;,
\end{equation}
which matches the case $\mu^f= 0$ (average projection) at $T= 0$\,.
On the other hand, the energy scale $\omega_0$ of spin excitations,
Eq.(\ref{eqn-enscaleresult}), is exponentially small compared to
$T$\,, therefore the opposite limit should apply,
\begin{equation}
  \label{eqn-flimit-high}
  T\gg\omega\,: \quad
  f(\omega - i\frac{\pi}{2}T)
    = \frac{1}{2}(1 + i)
      + \mathcal{O}(\frac{\omega}{T})\;\;,
\end{equation}
\end{subequations}
adding a significant imaginary part to the fermion spectral-function
$\widehat{G}(\omega)$\,. It requires a solution of the set of
Eqs.(\ref{eqn-nflex}), however, to reveal that $\widehat{G}(\omega)$
has no features\cite{note-fspecen} at $\omega\lesssim\omega_0$ (see
Fig.\ \ref{fig-fdos}). The fermion spectrum is governed by its
bandwidth $\sim J$\,, and therefore the crossover from
high-temperatures corresponding to Eq.(\ref{eqn-flimit-high}) to low
temperatures, where Eq.(\ref{eqn-flimit-low}) becomes valid, happens
at $T\sim J$\,. A more physical interpretation is provided in the
following Sections \ref{sec-const} and \ref{sec-sum}\,.

\section{Measuring the constraint}
\label{sec-const}

In order to understand the weak influence of the fermion constraint at
low temperature, it is useful to calculate the auxiliary-charge
fluctuations in average projection. Starting from
Eq.(\ref{eqn-expect-av}), we have to calculate
\begin{equation}
  \label{eqn-qfluct}
  \langle \Delta Q_i\, \Delta Q_j\rangle^\av
    = \langle Q_i Q_j\rangle^\av -
      \langle Q_i\rangle^\av\,\langle Q_j\rangle^\av\;.
\end{equation}
Since all charges $Q_i$ are conserved, ${}[\,Q_i\,,\, H\,{}]= 0$\,,
${}[\,Q_i\,,\, Q_j\,{}]= 0$\,, the correlation function
(\ref{eqn-qfluct}) may be obtained from the charge propagator in
Matsubara space as
\begin{eqnarray}
  \chi^Q_{ij}(i\nu)
    & = &  \nonumber 
      \int_0^{1/T}\mathrm{d}\tau\,
      e^{i\nu\tau}
      \langle\mathcal{T}_\tau\{\Delta Q_i(\tau)\, \Delta Q_j(0)\}\rangle^\av
      \\
    & = &  \label{eqn-qmatsu}
      \frac{1}{T}\,\delta_{\nu, 0}\,
      \langle \Delta Q_i\, \Delta Q_j\rangle^\av\;.
\end{eqnarray}
$\chi^Q_{ij}(i\nu)$ is conveniently calculated with the
Feynman-diagram rules introduced in Sect.\ \ref{sec-simple}\,, using
the Hamiltonain (\ref{eqn-allgrand}) with $\mu^f= 0$ and the bare
fermion Green's function (\ref{eqn-baregf})\,.

\textrm{Free spins:} In Section \ref{sec-simple} we first discussed
the limit $J= 0$\,. In that case, $\chi^Q$ is given by a simple bubble
of bare fermion Green's functions,
\begin{eqnarray*}
  \chi^Q_{ij}(i\nu)
    & = & - T\sum_{i\omega}\textrm{Tr}^\sigma{}[\,
        \overline{G}^0_{ij}(i\nu + i\omega)\,
        \overline{G}^0_{ji}(i\omega)\,{}]
      \\
    & = & \frac{1}{2T}\delta_{i,j}\delta_{\nu,0}\;\;,
\end{eqnarray*}
that is,
\begin{equation}
  \label{eqn-qflu-free}
  J= 0\,: \quad
  \langle \Delta Q_i\, \Delta Q_j\rangle^\av
    = \delta_{i,j}\frac{1}{2}\;.
\end{equation}
As expected, the auxiliary-charge fluctuations are finite. Note that
$\chi^Q_{ij}(i\nu)$ is local ($\sim\delta_{i,j}$) and static,
($\sim\delta_{\nu,0}$) in accordance with Eq.(\ref{eqn-qmatsu}), i.e.,
the conservation of the $Q_i$\,.

\emph{Mean-field theory:} The second example presented in Section
\ref{sec-simple} is the Hartree approximation. $\chi^Q$ is again given
by the simple bubble\cite{note-qfluct-hartree}, with
$\overline{G}_{ij}^0$ replaced by
\begin{displaymath}
  \overline{G}_{ij}(i\omega)
    = \delta_{i,j}{}[\,i\omega + (-1)^i \sigma^z\,h\,{}]^{-1}\;\;,
\end{displaymath}
with the Weiss field $h$ given in Eq.(\ref{eqn-selfhartree})\,. For
$T\ll J$ we obtain
\begin{equation}
  \label{eqn-qflu-hartree}
  \textrm{Hartree:} \quad
  \langle \Delta Q_i\, \Delta Q_j\rangle^\av
    = \delta_{i,j}2\exp(-h/T)\;\;,
\end{equation}
with $h= \frac{zJ}{4} + \mathcal{O}(e^{-J/T})$\,. That is, in the
magnetically ordered phase the unphysical charge fluctuations are
strongly suppressed, since the fermions develop a charge gap similar
to a spin-density-wave state.

\emph{Self-consistent theory:} The approximation discussed in Sections
\ref{sec-advanced} and \ref{sec-num} requires a more elaborate
calculation. For the case of average constraint, $\mu^f= 0$\,, the
approximation given by Eqs.(\ref{eqn-sflex}) and Fig.\ \ref{fig-flex}
has been studied earlier in Ref.\ \onlinecite{briwol04}\,. In Section
IV\,C of that paper the conserving-approximation scheme has been
applied to the spin susceptibility, leading to a vertex function in
the fermion bubble, which is determined by a Bethe--Salpeter equation.
The corresponding diagrams are shown in Fig.\ 6 of Ref.\
\onlinecite{briwol04}\,. For the auxiliary-fermion-charge
susceptibility we want to calculate, the diagrams are exactly the same,
except that the two spin vertices appearing in the bubble
$\widehat{\Pi}$ in Fig.\ 6 are to be replaced by charge vertices
$\sigma^0= 1$\,. The response function (\ref{eqn-qmatsu}) then reads
in wave-vector space
\begin{displaymath}
  \chi^Q(\mathbf{q}, 0)
    = -T\sum_{i\omega}\textrm{Tr}^{\sigma}{}[\,
      \sigma^0\,\overline{G}(i\omega)\,\overline{\Gamma}(\mathbf{q},
      i\omega)\,\overline{G}(i\omega)\,{}]\;.
\end{displaymath}
For $\langle \Delta Q\,\Delta Q\rangle^\av$ merely the static limit
$\displaystyle\lim_{\nu\to 0}\chi^Q(\mathbf{q}, i\nu)$ is needed.
With spin-rotation symmetry one has $\overline{G}= \sigma^0\,G$\,, and
it turns out that $\overline{\Gamma}= \sigma^0\,\Gamma$ (i.e., the
charge and spin channels do not mix in the vertex function), leading
to
\begin{equation}
  \label{eqn-flex-qbub}
  \chi^Q(\mathbf{q}, 0)
    = -2T\sum_{i\omega} G(i\omega)^2\,\Gamma(\mathbf{q}, i\omega)\;.
\end{equation}
The vertex function is specified through the following Bethe--Salpeter
equation, taken from the diagrams in Fig.\ 6 of Ref.\
\onlinecite{briwol04}\,,
\begin{equation}  \label{eqn-flex-bethe}
  \begin{array}[c]{l}
    \displaystyle 
    \Gamma(\mathbf{q}, i\omega) = 1 +
      \\[1ex]  \displaystyle 
    \mbox{}+
    \frac{3}{4}T\sum_{i\omega_1} G(i\omega_1)^2\,
      D(i\omega - i\omega_1)\,\Gamma(\mathbf{q}, i\omega_1) -
      \\[3ex]  \displaystyle 
    \mbox{}-
      \frac{3}{8}T\sum_{i\nu} G(i\omega + i\nu)
      \frac{1}{N_L}\sum_{\mathbf{k}}
      D(\mathbf{k} + \mathbf{q}, i\nu)\,D(\mathbf{k}, i\nu) \times
      \\[3ex]  \displaystyle 
    \mbox{}\times
      T\sum_{i\omega_1} G(i\omega_1)^2\,
      {}[\,G(i\omega_1 + i\nu) + 
      G(i\omega_1 - i\nu)\,{}]\,
      \Gamma(\mathbf{q}, i\omega_1)
  \end{array}
\end{equation}
The fermion Green's function $G(i\omega)$ and the local spin-spin
interaction $D(i\nu)$ have to be taken from the solution of the
Eqs.(\ref{eqn-sflex}) for $\mu^f= 0$\,. The non-local spin interaction
appearing in (\ref{eqn-flex-bethe}) reads
\begin{displaymath}
  D(\mathbf{q}, i\nu)
    = - J(\mathbf{q}) + J(\mathbf{q})^2\,\chi(\mathbf{q}, i\nu)\;\;,
\end{displaymath}
with $\chi$ from Eq.(\ref{eqn-sflex-chi})\,.

In Appendix \ref{sec-app-qbub} it is shown that the 2nd term in
Eq.(\ref{eqn-flex-bethe}) actually becomes zero by symmetry arguments,
i.e., the Bethe--Salpeter equation simplifies to
\begin{equation}  \label{eqn-flex-bethe2}
  \Gamma(\mathbf{q}, i\omega)
    = 1 + \frac{3}{4}T\sum_{i\omega_1} G(i\omega_1)^2\,
  D(i\omega - i\omega_1)\,\Gamma(\mathbf{q}, i\omega_1)
\end{equation}
Instead of solving the last equation numerically, we find it more
instructive to aim at an approximate analytical solution. We employ
the static approximation introduced in Ref.\ \onlinecite{briwol04}\,,
i.e., let $D(i\nu)= D(0)\delta_{\nu, 0}$ in Eq.(\ref{eqn-flex-bethe2})
as well as the fermion self-energy (\ref{eqn-sflex-self})\,. The
calculation closely follows Sections IV\,A and C of Ref.\
\onlinecite{briwol04}\,, leading to
\begin{displaymath}
  \Gamma(\mathbf{q}, i\omega)
    = \frac{1}{1 - (\omega_f/2)^2 G(i\omega)^2}
      \;\;,\;\;\;
  \omega_f
    = J\frac{16}{3\pi} + \mathcal{O}(T^2)
\end{displaymath}
for temperatures $T\ll J$\,. $\omega_f$ is a typical 1/2 bandwidth of
the continuous fermion spectrum, compare the bottom panel of Fig.\
\ref{fig-fdos}\,. Performing the Matsubara-sum in
Eq.(\ref{eqn-flex-qbub}) eventually leads to
\begin{displaymath}
  \chi^Q(\mathbf{q}, 0)
    = \frac{1}{\omega_f}\Phi^Q(T/J)
      \;\;,\;\;\;
  \Phi^Q(t)
    = \frac{4}{\pi} + \mathcal{O}(t^2)\;.
\end{displaymath}
If the vertex function is ignored, $\Gamma \to 1$\,, the result does
not change significantly, $\Phi^Q(t) \to \frac{16}{3\pi} +
\mathcal{O}(t^2)$\,. Note that $\chi^Q$ is independent of
$\mathbf{q}$\,, i.e., local.

From Eq.(\ref{eqn-qmatsu}) we thus find the auxiliary-charge
fluctuations of our self-consistent approximation in average
projection,
\begin{equation}
  \label{eqn-qflu-flex}
  \textrm{self cons.:} \quad
  \langle \Delta Q_i \,\Delta Q_j \rangle^\av
    = \delta_{i, j}
      \frac{3 T}{4 J} + \mathcal{O}(T^3)\;.
\end{equation}
Since the fermion spectrum is gapless\cite{note-fspecen} around
$\omega= 0$ (see Fig.\ \ref{fig-fdos}), and the vertex function has
very little effect, $\chi^Q$ comes out Pauli-like. The explicit $T$
factor in Eq.(\ref{eqn-qmatsu}) suppresses the charge fluctuations at
low temperature. Compared to the magnetically ordered state described
in mean-field (Hartree) theory, where the fermions develop a gap $\sim
J$ (see Eq.(\ref{eqn-qflu-hartree})), the suppression of charge
fluctuations is much weaker\cite{note-pregap}.  However, at $T\to 0$
the unphysical fluctuations still vanish, and the average projection
becomes exact.

\section{Summary and conclusions}
\label{sec-sum}

Resummed perturbation theory is a powerful tool for calculating
dynamical properties of strongly correlated electron systems. In this
paper we focused on the spin-1/2 antiferromagnetic quantum Heisenberg
model on the two dimensional square lattice. Summing infinite classes
of contributions in perturbation theory is most economically done
within a quantum-field-theoretic formulation employing canonical
fields. We therefore use an auxiliary-particle representation of spin
operators, which is a faithful representation in the physical sector
of the Hilbert space. 

The use of auxiliary fermions requires a projection onto the physical
part of the fermion-Fock space, where the fermion charge $Q_i$ equals
one for each lattice site.

While for a single lattice site the projection may be done exactly,
e.g., by introducing an auxiliary-fermion energy $\lambda$\,, which is
sent to infinity at the end of the calculation\cite{abr65,col84},
these standard methods cannot directly be generalized to effect the
projection at each lattice site independently (this would require
handling a large number of independent limiting procedures, an
impossible task in practice).

For lattice systems the most simple approach to the projection is an
approximative treatment, where a global chemical potential (Lagrange
multiplier) $\mu^f$ is introduced, which is sufficient to fulfill the
constraint on the thermal average, $\langle Q_i\rangle= 1$\,.

However, Popov and Fedotov have proposed a rather unusual projection
method, where a global \emph{imaginary-valued} chemical potential
$\mu^f= i\frac{\pi}{2}T$ leads to an \emph{exact} cancellation of
unphysical states, therefore enforcing the operator constraint $Q_i=
1$\,.  Unfortunately, the Popov-Fedotov method may not
straight-forwardly be generalized to systems away from particle-hole
symmetry\cite{grosjohn90}.

In this paper we explored the usability of this concept by identifying
the conditions to be satisfied by any, necessarily self-consistent,
approximation scheme. Most important is the conservation of the
fermion charge $Q_i$ by the model Hamiltonian, ${}[Q_i, H{}]= 0$\,. If
the approximation under consideration violates this conservation law,
results become meaningless (see Sec.\ \ref{sec-intro} and App.\
\ref{sec-app-greens})\,. Therefore, self-consistent approximations are
most safely based on the conserving-approximation principle. Any
hopping of auxiliary fermions, for example, is precluded by this
requirement: The fermions are strictly local entities. The physically
observable momentum dependence of spin correlation functions
originates from the momentum dependence of the exchange interaction.

Within the Popov--Fedotov approach the well known
Feynman-skeleton-diagram expansion is applicable in conjunction with
an exact projection of the auxilary particles onto the physical
Hilbert space. We have shown in some detail how a self-consistent
approximation, which goes far beyond mean-field theory similar to the
``fluctuation-exchange approximation'', can be formulated using
complex-valued spectral functions of the (unphysical) renormalized
fermion propagator. The resulting equations have been solved by
numerical iteration.

We applied the Popov-Fedotov method on several approximation levels:
the free spin, the Hartree approximation (magnetic mean-field theory),
and the above-mentioned self-consistent approximation, using both
\emph{average projection} ($\mu^f= 0$) and \emph{exact projection}
($\mu^f= i\frac{\pi}{2}T$)\,.  The results obtained for the latter
approximation show the expected suppression of the ordered state down
to zero temperature, the exponential divergence of the spin
correlation length, and a spin-structure factor consistent to the
dynamical scaling hypothesis.

A comparison of the results from average and exact projection reveals
that there is a significant effect of the exact projection at higher
temperatures. In the limit of low temperature, however, the deviation
of the average-constraint results from the exact-constraint results
become (numerically) indistinguishable, except for the case $J= 0$
(free spins).

In order to support this observation, we calculated the fluctuations
$\langle \Delta Q_i\,\Delta Q_j\rangle$ of the auxiliary-fermion
charge within the average-projection scheme. We find (by analytical
calculation) $\;\lim_{T\to 0}\langle \Delta Q_i\,\Delta Q_j\rangle=
0$\,, except for the case of free spins, where $\langle \Delta
Q_i\,\Delta Q_j\rangle$ stays finite as $T\to 0$\,. That is, as long
as the spin--spin interaction $J$ is taken into account, the
fermion-charge fluctuations into unphysical Hilbert-space states are
quenched at $T= 0$\,. If temperature is increased from zero, we find
that $\langle \Delta Q_i\,\Delta Q_j\rangle$ raises continuously with
$T$\,.

These at first sight surprising results find their explanation in the
tendency towards antiferromagnetic order in the interacting system,
which helps to suppress the fluctuations in the fermion-occupation
number: Starting from the physical (``true'') ground state, which
features long-range magnetic order\cite{man91}, a fluctuation of the
fermion charge $Q_l=1$ at some site $l$ into an unphysical
state\cite{note-qtransfer} with $Q_l= 0$ or $Q_l= 2$ is equivalent to
\emph{removing the spin} $\mathbf{S}_l$ in the Hamiltonian (recall
Eqs.(\ref{eqn-destruct}) and (\ref{eqn-equalen}))\,. The lowest-lying
state in this unphysical subspace thus lacks the binding energy of the
spin at site $l$\,, which is of order $J$\,. Therefore, the ground
state in the Fock space of arbitrary fermion occupancy is the ``true''
ground state in the physical segment, and the lowest-lying unphysical
state is separated from the ground state by a gap\cite{note-qflexpon}
$\Delta E_Q\sim J$\,.

Consequently, at low temperatures $T\ll \Delta E_Q$\,, to a good
approximation the exact projection may be omitted in favour of the
technically somewhat simpler average-projection approach. At $T= 0$
the approximate treatment of the constraint even becomes exact. Note
that $T\ll \Delta E_Q$ does not impose any restriction on the
excitation energy $\omega$\,, e.g., in the structure factor
$S(\mathbf{q}, \omega)$\,: Since the fermion charge is conserved
locally, all excitations at any $\omega$ out of the ground state
remain in the physical Hilbert space.

The above argument is quite apparent for magnetically ordered systems.
However, it should also apply to systems without magnetic order but
strong correlations in the ground state. Examples are the various
valence bond states discussed for, e.g., Heisenberg models with
frustration\cite{mislhu05}. In these systems the gap $\Delta E_Q$ to
unphysical states is also expected to be $\sim J$\,. Somewhat
different examples are systems with a ground state that is dominated
by local Kondo singlets. Here the gap $\Delta E_Q$ is exponentially
small in $J$\,, since the binding energy of a localized spin to the
Fermi sea is given by the exp.\ small Kondo energy $T_K$\,. For
calculations in the important temperature range $T\gtrsim T_K$ a solid
treatment of the fermion constraint is therefore desirable.

As far as the low-temperature behavior is concerned, the criticism of
the auxiliary-particle approach often expressed in view of the
uncontrolled handling of the constraint may be refuted on the basis of
the results presented here. However, one has to keep in mind that the
above arguments are based on the assumption that the approximation
method (whether based on self-consistent diagrams or functional
integrals) does conserve the local fermion charge $Q_i$\,.

The Popov--Fedotov approach opens the way to using resummed
perturbation theory in specific strongly correlated systems, on the
basis of standard Feynman diagrams, and for all temperatures. It
requires identifying and performing the summation of physically
relevant terms (diagram classes), which, however, remains a challenge
for these systems. The self-consistent approximation presented here,
for example, still fails to satisfy the notoriously hard to meet sum
rule on the local spin moment. More elaborate resummation schemes are
necessary to correct this and other deficiencies, the reward being a
detailed description of the spin dynamics not accessible by any other
analytical method.

\section{Acknowledgements}

We acknowledge useful discussions with J.~Reuther. This work has
partially been supported by the Research Unit ``Quantum-Phase
Transitions'' of the Deutsche Forschungsgemeinschaft.

\appendix

\section{Properties of Green's functions in the Popov--Fedotov
  technique}
\label{sec-app-greens}

In this appendix we consider thermal (Matsubara) Green's functions in
the Popov--Fedotov scheme. For some operators $A$ and $B$\,, which are
both either fermionic ($s= +1$) or bosonic ($s= -1$), the Green's
function is defined as\cite{negorl},
\begin{displaymath}
  G(\tau)
    = (-s)\langle \mathcal{T}_\tau\{ A(\tau) B(0)\}\rangle^\ppv
      \;\;,\;\;\;
  -\beta\le \tau< \beta\;\;,
\end{displaymath}
with $\beta= 1/T$\,, the thermal expectation value as defined in
Eq.(\ref{eqn-expect-ppv}) and (\ref{eqn-zppdef}), the Hamiltonian
given by Eqs.(\ref{eqn-grand}), (\ref{eqn-hbergham}), and the usual
``time''-ordering symbol
\begin{displaymath}
  \mathcal{T}_\tau\{A(\tau) B(0)\}
    = A(\tau) B(0) \Theta(\tau) -s\,
      B(0) A(\tau) \Theta(-\tau)\;.
\end{displaymath}
The fact that $H^\ppv$\,, Eq.(\ref{eqn-grand}), is non-Hermitian, does
not influence the (anti-) symmetry properties resulting from the
cyclic invariance of the trace. Therefore it is sufficient to consider
$\tau> 0$\,, i.e.,
\begin{displaymath}
  G(\tau)
    = (-s)\frac{1}{Z^\ppv}
      \textrm{Tr}^f{}[\,
        e^{-(\beta - \tau) H^\ppv} A
        e^{-\tau H^\ppv} B\,{}]\;.
\end{displaymath}
Using Eqs.(\ref{eqn-sgl}) and (\ref{eqn-grand}) this
becomes
\begin{eqnarray}
  G(\tau)
    & = & \nonumber
      (-s)\frac{1}{Z^\ppv}
      \sum_{c_Q, n_Q}\sum_{c'_Q, n'_Q} \cdot
      \\
    &   & \nonumber \cdot
      \langle c_Q, n_Q|A|c'_Q, n'_Q\rangle
      \langle c'_Q, n'_Q|B|c_Q, n_Q\rangle
      \cdot
      \\
    &   & \nonumber \cdot
      e^{-\beta E(c_Q, n_Q)}\,
      e^{\tau{}[\,E(c_Q, n_Q) - E(c'_Q, n'_Q)\,{}]}
      \cdot
      \\
    &   & \cdot
    \label{eqn-green-spec1}
      \big(\prod_{k= 1}^{N_L}e^{i\frac{\pi}{2}Q_k}\big)\,
      \big(\prod_{q= 1}^{N_L}
        e^{-i\frac{\pi\tau}{2\beta}{}[\,Q_q - Q'_q\,{}]}\big)
\end{eqnarray}
$Q_q$ and $Q'_q$ denote the auxiliary charge on lattice site $q$ as it
appears in the charge configurations $c_Q$ and $c'_Q$\,, respectively.

\emph{Physical propagator:} The simplest physical Green's function is
the dynamical spin susceptibility (\ref{eqn-sus-def}),
(\ref{eqn-sus-ppv}), for two lattice sites $l$ and $m$\,,
\begin{displaymath}
  \chi_{lm}^{\mu\bar{\mu}}(\tau)
    = \langle \mathcal{T}_\tau\{
        S^\mu_l(\tau) S^{\bar{\mu}}_m(0)
        \rangle^\ppv\;.
\end{displaymath}
Due to the property (\ref{eqn-destruct}) of spin operators, the
fermion charge on the sites $l, m$ is automatically constrained to 1
in Eq.(\ref{eqn-green-spec1}), i.e., $Q_l= Q_m= Q'_l= Q'_m= 1$\,. For
all other sites $p\ne l, m$\,, the orthonormal matrix elements in
Eq.(\ref{eqn-green-spec1}) lead to $Q_p= Q'_p$\,, thus we have $c'_Q=
c_Q$\,, and the second factor $(\prod_{q= 1}^{N_L}e^{\cdots})$ becomes
1\,.  Now the term $(\prod_{k= 1}^{N_L}\exp(i\frac{\pi}{2}Q_k))$\,, in
combination with the property (\ref{eqn-equalen}) of the energies,
leads to a cancellation of all unphysical states with charge $Q_p= 0$
and $Q_p= 2$ on any lattice site $p\ne l, m$\,. Thus, only physical
states with a single fermion per site, $c_Q= c'_Q= c_Q^{phys}= (1, 1,
\ldots, 1)$ remain in Eq.(\ref{eqn-green-spec1})\,. Using the notation
\begin{displaymath}
  E_n= E(c_Q^{phys}, n_Q)  \;\;,\;\;\;
  |n\rangle= |c_Q^{phys}, n_Q\rangle
\end{displaymath}
for energies and states in the physical subspace,
Eq.(\ref{eqn-green-spec1}) reads
\begin{equation}
  \label{eqn-green-phys}
  \chi_{lm}^{\mu\bar{\mu}}(\tau)
    = \frac{(i)^{N_L}}{Z^\ppv}
      \sum_{n, n'}
      \langle n|S^\mu_l|n'\rangle
      \langle n'|S^{\bar{\mu}}_m|n\rangle
      e^{-\beta E_n}
      e^{\tau(E_n - E_{n'})}
\end{equation}
With the result (\ref{eqn-physpart}) for the partition function, this
is exactly the expression we would have obtained directly, working in
the physical Hilbert space.

\emph{Green's function of the fermions:} The fermion propagator is not
a meaningful physical quantity. However, within a self-consistent
diagrammatic expansion of, e.g., the dynamical spin susceptibility,
the renormalized fermion Green's function is of technical importance.
Therefore it is useful to derive some exact properties of the Green's
function
\begin{equation}
  \label{eqn-green-fgfdef}
  G_{lm,\sigma}(\tau)
    = - \langle\mathcal{T}_\tau\{
          f_{l \sigma}(\tau) f^\dagger_{m \sigma}(0)\}
          \rangle^\ppv\;.
\end{equation}
The matrix elements in Eq.(\ref{eqn-green-spec1}) now read
\begin{displaymath}
  \langle c_Q, n_Q|f_{l \sigma}|c'_Q, n'_Q\rangle
  \langle c'_Q, n'_Q|f^\dagger_{m \sigma}|c_Q, n_Q\rangle\;.
\end{displaymath}
$f^\dagger_{m \sigma}$ increases the auxiliary charge at lattice site
$m$ by 1, which can only be compensated by $f_{l \sigma}$\,, that is,
the exact fermion propagator is local, $l= m$\,. For all sites $p\ne
l$ the arguments from above hold: The orthonormal wave functions lead
to $Q_p= Q'_p$\,, and the unphysical constributions with $Q_p= 0, 2$
cancel. The states that remain in the trace in
Eq.(\ref{eqn-green-spec1}) then have $Q_p= 1$ at all sites $p\ne l$
and some charge $Q_l, Q'_l\in \{0, 1, 2\}$ at site $l$\,. With the
notation
\begin{displaymath}
  |Q, n\rangle^l= |c_q, n_Q\rangle
  \quad\textrm{for}\quad
  Q_l= Q \;\;,\;\;\;
  Q_p= 1\,, \;p\ne l\;\;,
\end{displaymath}
and similarly $E^l(Q, n)$ for the eigenenergies, the Green's function
(\ref{eqn-green-spec1}), (\ref{eqn-green-fgfdef}) becomes
\begin{eqnarray}  \label{eqn-green-gferm}
  G_{lm, \sigma}(\tau)
    & = & (-1)\delta_{lm}\frac{(i)^{N_L - 1}}{Z^\ppv}
      \sum_{Q= 0, 1}
      \sum_{n, n'}
      \cdot
      \\
    &   & \mbox{}  \nonumber
      \cdot
      |\,{}^l\!\langle Q, n|f_{l \sigma}|Q+1, n'\rangle^l|^2
      \cdot
      \\
    &   & \mbox{}  \nonumber
      \cdot
      \exp\big({-\beta E^l(Q, n)}\big)\,
      \exp\big({i\frac{\pi}{2}Q}\big)
      \cdot
      \\
    &   & \mbox{}  \nonumber
      \cdot
      \exp\big({\tau{}\big[\,E^l(Q, n) - E^l(Q + 1, n') 
        + i\frac{\pi}{2\beta}\,{}\big]}\big)
\end{eqnarray}
The energies and states that occur in Eq.(\ref{eqn-green-gferm}) are
now denoted by
\begin{eqnarray*}
  Q= 1\,: &  &
    E^l(1, n) = E_n\;\;,
    \\
  Q= 0, 2\,:  &  &
    E^l(0, n') = E^l(2, n') = E^l_{n'}\;\;,
\end{eqnarray*}
and
\begin{eqnarray*}
  Q= 1\,: &  &
    |1, n\rangle^l = |n\rangle\;\;,
    \\
  Q= 0\,: &  &
    f^\dagger_{l \sigma}|0, n'\rangle^l = |\sigma, n'\rangle^l\;\;,
    \\
  Q= 2\,: &  &
    f_{l \sigma}|2, n'\rangle^l = \pm|-\sigma, n'\rangle^l\;.
\end{eqnarray*}
According to Eqs.(\ref{eqn-physconfig}), (\ref{eqn-sgl}) the $E_n$ and
$|n\rangle$ are the eigenenergies and -states of the model Hamiltonian
$H$\,, Eq.(\ref{eqn-hbergham})\,. Referring to
Eqs.(\ref{eqn-destruct}) and (\ref{eqn-equalen}), the $E_{n'}^l$ can
be interpreted as the eigenenergies of $H$ \emph{with a ``defect'' at
  site $l$}, i.e., with all couplings $J$ to the spin at site $l$ set
to zero. The states $|\sigma, n'\rangle^l$ therefore contain the
orientation $\sigma\in\{\uparrow, \downarrow\}$ of the resulting free
spin at site $l$ as a good quntum number. The set of quantum numbers
$n'$ as well as the $E_{n'}^l$ do not depend on $\sigma$\,.  Note that
the number of states is the same, $\#\{|n\rangle\}= \#\{|\sigma,
n'\rangle\}= 2^{N_L}$\,.

The fermion Green's function (\ref{eqn-green-spec1}) now reads,
\begin{eqnarray}
  \label{eqn-green-gtau}
  G_{l, \sigma}(\tau)
    & = & \frac{-1}{Z}
      \sum_{n, n'}\cdot
      \\
    &   & \nonumber
      \mbox{}\cdot\bigg\{
      e^{-\beta E_n}
      |\langle n| -\sigma, n'\rangle^l |^2
      e^{\tau{}[\,E_n - E^l_{n'} + i\frac{\pi}{2\beta}\,{}]} -
      \\
    &   & \nonumber
      \mbox{} -
      i\,e^{-\beta E^l_{n'}}
      |\langle n| \sigma, n'\rangle^l |^2
      e^{\tau{}[\,E^l_{n'} - E_n + i\frac{\pi}{2\beta}\,{}]}
      \bigg\}
\end{eqnarray}

In frequency space, the fermion propagator is given by
\begin{equation}  \label{eqn-green-mats}
  G_{lm, \sigma}(i\omega)
    = \int_0^\beta\mathrm{d}\tau\,
      e^{i\omega \tau}
      G_{lm, \sigma}(\tau)\;\;,
\end{equation}
with the fermionic (odd) Matsubare frequency $\omega= (2n +
1)\frac{\pi}{\beta}$\,. Inserting Eq.(\ref{eqn-green-gtau}) into
Eq.(\ref{eqn-green-mats}) and utilizing the earlier result
Eq.(\ref{eqn-physpart}) for $Z^\ppv$\,, we obtain the fermion Green's
function,
\begin{equation}
  \label{eqn-green-fgfspec}
  G_{lm, \sigma}(i\omega)
    = \delta_{lm}
      \int_{-\infty}^\infty\mathrm{d}\varepsilon\,
      \frac{\widehat{G}_{l, \sigma}(\varepsilon)}
           {i\omega + i\frac{\pi}{2\beta} - \varepsilon}\;\;,
\end{equation}
with the complex-valued spectral function
\begin{equation}
  \label{eqn-green-cmplxspec}
  \widehat{G}_{l, \sigma}(\varepsilon)
    = \rho^1_{l, \sigma}(\varepsilon) +
      i\,\rho^2_{l, \sigma}(\varepsilon)\;\;,
\end{equation}
\begin{eqnarray*}
  \rho^1_{l, \sigma}(\varepsilon)
    & = & \frac{1}{Z}
      \sum_{n, n'} e^{-\beta E_n}\cdot
      \\
    &   & \cdot\bigg\{
      |\langle n|\sigma, n'\rangle^l|^2\,
      \delta\big(\varepsilon - {}[\,E_n - E^l_{n'}{}]\big) +
      \\
    &   & \mbox{} +
      |\langle n|-\sigma, n'\rangle^l|^2\,
      \delta\big(\varepsilon - {}[\,E^l_{n'} - E_n{}]\big)
      \bigg\}
\end{eqnarray*}
\begin{eqnarray*}
  \rho^2_{l, \sigma}(\varepsilon)
    & = & \frac{1}{Z}
      \sum_{n, n'} e^{-\beta E^l_{n'}}\cdot
      \\
    &   & \cdot\bigg\{
      |\langle n|-\sigma, n'\rangle^l|^2\,
      \delta\big(\varepsilon - {}[\,E^l_{n'} - E_n{}]\big) -
      \\
    &   & \mbox{} -
      |\langle n|\sigma, n'\rangle^l|^2\,
      \delta\big(\varepsilon - {}[\,E_n - E^l_{n'}{}]\big)
      \bigg\}
\end{eqnarray*}

\emph{Sum rule and symmetry:} The $|n\rangle$ as well as the $|\sigma,
n'\rangle^l$ form a complete normalized basis in the physical Hilbert
space,
\begin{displaymath}
  \sum_n|n\rangle\langle n|= 1  \;\;,\;\;\;
  \sum_{n', \sigma}
    {}^l\!|\sigma, n'\rangle\langle \sigma, n'|^l
    = 1\;\;,
\end{displaymath}
and therefore integrating Eqs.(\ref{eqn-green-cmplxspec}) over
$\varepsilon$ leads to the sum rule
\begin{displaymath}
  \int_{-\infty}^\infty\mathrm{d}\varepsilon\,
  \rho^1_{l, \sigma}(\varepsilon)
    = 1
    \;\;,\;\;\;
  \int_{-\infty}^\infty\mathrm{d}\varepsilon\,
  \rho^2_{l, \sigma}(\varepsilon)
    = 0\;\;,
\end{displaymath}
\begin{equation}
  \label{eqn-green-sum}
  \;\;\;\Rightarrow\;\;\;
  \int_{-\infty}^\infty\mathrm{d}\varepsilon\,
  \widehat{G}_{l, \sigma}(\varepsilon)
    = 1\;.
\end{equation}

From Eq.(\ref{eqn-green-cmplxspec}) we can also read off a
``particle--hole'' symmetry,
\begin{displaymath}
  \rho^1_{l, \uparrow}(\varepsilon)
    = \rho^1_{l, \downarrow}(-\varepsilon)
    \;\;,\;\;\;
  \rho^2_{l, \uparrow}(\varepsilon)
    = - \rho^2_{l, \downarrow}(-\varepsilon)\;\;,
\end{displaymath}
\begin{equation}
  \label{eqn-green-phsym}
  \;\;\;\Rightarrow\;\;\; 
  \widehat{G}_{l, \uparrow}(\varepsilon)
    = \widehat{G}_{l, \downarrow}(- \varepsilon)^\ast\;.
\end{equation}
In the paramagnetic phase, where the overlaps in
Eq.(\ref{eqn-green-cmplxspec}) are spin degenerate,
\begin{displaymath}
  \langle n|\uparrow, n'\rangle^l  =
  \langle n|\downarrow, n'\rangle^l\;\;,
\end{displaymath}
Eq.(\ref{eqn-green-phsym}) simplyfies to the result already quoted in
Eq.(\ref{eqn-phsymm})\,.

\emph{The expectation value $\langle Q_l\rangle$\,:} In order to
conclude this appendix, we discuss the average auxiliary charge
$\langle Q_l\rangle$ at site $l$\,.

In the enlarged Hilbert space the expectation value $\langle
Q_l\rangle^\ppv$ can be formally calculated; however, although $Q_l$
is a gauge-invariant operator, it does not fulfill the property
(\ref{eqn-destruct}) of physical observables, and therefore the result
becomes meaningless.  This is most easily demonstrated by explicitly
calculating $\langle Q_l\rangle^\ppv$\,:

Using the Green's function (\ref{eqn-green-fgfdef}) it may be written
as
\begin{displaymath}
  \langle Q_l\rangle^\ppv
    = \sum_{\sigma}
      G_{l, \sigma}(\tau= 0_-)\;.
\end{displaymath}
In Eq.(\ref{eqn-green-gtau}) the propagator has been given for $\tau>
0$\,, which can be utilized by help of the anti-symmetric property of
fermionic Green's functions,
\begin{displaymath}
  \tau< 0\,: \quad
  G(\tau)= - G(\tau + \beta)\;.
\end{displaymath}
Thus we find from Eq.(\ref{eqn-green-gtau}), setting $\tau=
\beta$\,,
\begin{eqnarray*}
  \langle Q_l\rangle^\ppv
    & = & \frac{1}{Z}\sum_n e^{- \beta E_n}\langle n|n\rangle +
      \\
    &   & \mbox{} +
      i\,\frac{1}{Z}\sum_{n'}\sum_{\sigma}
      e^{-\beta E^l_{n'}}
      \,{}^l\!\langle \sigma, n'|\sigma, n'\rangle^l
      \\
    & = & \frac{1}{Z}(Z + i Z^l)
\end{eqnarray*}
Here $Z$ is the partition function of $H$\,,
Eq.(\ref{eqn-hbergham})\,, in the physical subspace, while $Z^l$ is
the partition function of $H$ with the ``defect'' at site $l$\,, i.e.,
with all couplings $J$ to the site $l$ made zero. Since all
interactions in $H$ are short ranged, $Z$ and $Z^l$ become equal in
the thermodynamic limit,
\begin{equation}
  \label{eqn-green-qexpect}
  N_L\to\infty\,: \quad
  \langle Q_l\rangle^\ppv
    = (1 + i\frac{Z^l}{Z})
    \;\to\; (1 + i)
\end{equation}
Note that $Z$ and $Z^l$ contain the same number of states,
$(2)^{N_L}$\,.

\section{Derivation of the self-consistent Equations
  (\ref{eqn-nflex})}
\label{sec-app-deriv}

In this appendix the intermediate steps in going from
Eqs.(\ref{eqn-sflex}) to Eqs.(\ref{eqn-nflex}) are presented.

Starting point is the spectral representation (\ref{eqn-gfspect}) or
(\ref{eqn-green-fgfspec}) of the fermion Green's function
(\ref{eqn-sflex-dyson})\,. The Matsubara frequency $i\omega$ can be
analytically continued to the complex plane, $i\omega\to z$\,, with
$G(z)$ showing a branch cut at $\textrm{Im}(z)=
-\frac{\pi}{2\beta}$\,. Close to this cut, at $z= \omega -
i\frac{\pi}{2\beta} \mp i0_+$\,, ($0_+$ is a positive infinitesimal)
we have
\begin{equation}
  \label{eqn-deriv-fgfdecomp}
  G(\omega - i\frac{\pi}{2\beta} \mp i0_+)
    = \pm i\pi \widehat{G}(\omega) + \overline{G}(\omega)\;\;,
\end{equation}
with the spectral function $\widehat{G}$ and its Hilbert transform
\begin{displaymath}
  \overline{G}(\omega)
    = \textrm{P}\!\!
      \int_{-\infty}^\infty\mathrm{d}\varepsilon\,
      \frac{\widehat{G}(\varepsilon)}{\omega - \varepsilon}\;.
\end{displaymath}
For the susceptibilities $\Pi(i\nu)$ and $D(i\nu)$\,, appearing in
Eq.(\ref{eqn-sflex-bubble}) and (\ref{eqn-sflex-effint}), the usual
analytic continuation of the bosonic Matsubara frequency $i\nu$ to the
real axis applies, $i\nu\to \omega + i0_+$\,,
\begin{equation}
  \label{eqn-deriv-cont}
  \begin{array}[c]{rcl} \displaystyle 
  \Pi(i\nu)
    & \to &
      \Pi(\omega + i0_+) = \Pi'(\omega) + i\,\Pi''(\omega)\;\;,
      \\ \\ \displaystyle 
  D(i\nu)
    & \to &
      D(\omega + i0_+) = D'(\omega) + i\,D''(\omega)\;.
  \end{array}
\end{equation}
The imaginary part $D''(\omega)$ represents the spectral function of
the effective local interaction,
\begin{equation}
  \label{eqn-deriv-effint}
  D(i\nu)
    = \frac{1}{\pi}
      \int_{-\infty}^\infty\mathrm{d}\varepsilon\,
      \frac{D''(\varepsilon)}{\varepsilon - i\nu}\;.
\end{equation}
Note that $D''(\omega)$ obeys the symmetry
\begin{equation}
  \label{eqn-deriv-dsymm}
  D''(-\omega)= - D''(\omega)\;\;,
\end{equation}
which comes from $\chi(\mathbf{q}, \omega)= \chi^\ast(-\mathbf{q},
-\omega)$ in Eq.(\ref{eqn-sflex-effint})\,. For $\Pi(i\nu)$\,,
equations similar to (\ref{eqn-deriv-effint}) and
(\ref{eqn-deriv-dsymm}) hold.

The fermion self-energy (\ref{eqn-sflex-self}) is re-written using
the spectral representations (\ref{eqn-green-fgfspec}) and
(\ref{eqn-deriv-effint}),
\begin{eqnarray*}
  \Sigma(i\omega)
    & = & -\frac{3}{4\pi}
      \int\mathrm{d}\varepsilon\,
      D''(\varepsilon)\,
      \int\mathrm{d}\varepsilon'\,
      \widehat{G}(\varepsilon')\,\cdot
      \\
    &   & \mbox{}\quad\cdot
      \frac{1}{\beta}
      \sum_{i\nu}
      \frac{1}{(i\nu - \varepsilon)
        (i\nu + i\omega + i\frac{\pi}{2\beta} - \varepsilon')}
      \\
    & = & \frac{3}{4\pi}
      \int\mathrm{d}\varepsilon\,
      D''(\varepsilon)\,
      \int\mathrm{d}\varepsilon'\,
      \widehat{G}(\varepsilon')
      \frac{g(\varepsilon) + f(\varepsilon' - i\frac{\pi}{2\beta})}
           {i\omega + i\frac{\pi}{2\beta} + \varepsilon - \varepsilon'}
\end{eqnarray*}
$f$ and $g$ stand for the Fermi and Bose function. Apparently,
$\Sigma(i\omega)$ obeys a spectral representation similar to
Eq.(\ref{eqn-green-fgfspec}), namely,
\begin{equation}
  \label{eqn-deriv-selfspec}
  \Sigma(i\omega)
    = \int\mathrm{d}\varepsilon\,
      \frac{\widehat{\Sigma}(\varepsilon)}
           {i\omega + i\frac{\pi}{2\beta} - \varepsilon}\;\;,
\end{equation}
with the (complex valued) spectral function
\begin{displaymath}
  \widehat{\Sigma}(\omega)
    = \frac{3}{4\pi}\int\mathrm{d}\varepsilon\,
      D''(\varepsilon)
      \widehat{G}(\varepsilon + \omega)
      {}[\,g(\varepsilon) + f(\varepsilon + \omega - i\frac{\pi}{2\beta})\,{}]
\end{displaymath}
and its Hilbert transform $\overline{\Sigma}$\,, given in
Eq.(\ref{eqn-nflex-selfhil}) above.

Introducing the structure factor of the renormalized interaction,
\begin{displaymath}
  U(\omega)
    = {}[\,1 + g(\omega)\,{}] D''(\omega)
    \;\;,\;\;\; 
\end{displaymath}
which is by Eq.(\ref{eqn-deriv-dsymm}) equivalent to
\begin{displaymath}
  U(-\omega)
    = g(\omega) D''(\omega)
    \;\;,\;\;\; 
\end{displaymath}
and using the relation
\begin{displaymath}
  g + f
    = g(1 - f) + (1 + g)f
    \;\;,\;\;\; 
\end{displaymath}
the spectrum $\widehat{\Sigma}$ takes the form
Eq.(\ref{eqn-nflex-selfspec}), with the short hands $\widehat{G}^+$
and $\widehat{G}^-$ defined in Eqs.(\ref{eqn-nflex-gfplus}) and
(\ref{eqn-nflex-gfmin})\,.

The fermion spectrum $\widehat{G}$ is obtained from the Dyson's
equation (\ref{eqn-sflex-dyson}) using Eq.(\ref{eqn-deriv-fgfdecomp}),
i.e.,
\begin{eqnarray*}
  \widehat{G}(\omega)
    & = & \frac{1}{2\pi i}\left[\,
        \frac{1}{\omega - \Sigma(\omega - i\frac{\pi}{2\beta} - i0_+)}
        \right. -
      \\
    &   & \mbox{}- \left.
        \frac{1}{\omega - \Sigma(\omega - i\frac{\pi}{2\beta} + i0_+)}
        \,\right]
\end{eqnarray*}
By inserting the decomposition
\begin{displaymath}
  \Sigma(\omega - i\frac{\pi}{2\beta} \mp i0_+)
    = \pm i\pi \widehat{\Sigma}(\omega) + \overline{\Sigma}(\omega)\;\;,
\end{displaymath}
which results from the spectral representation
(\ref{eqn-deriv-selfspec}) and Eq.(\ref{eqn-nflex-selfhil}), we obtain
$\widehat{G}$ as given in Eq.(\ref{eqn-nflex-gf})\,.

In the fermion bubble $\Pi(i\nu)$\,, Eq.(\ref{eqn-sflex-bubble}), the
spectral representation (\ref{eqn-green-fgfspec}) of the fermion
Green's function is inserted, and we arrive at
\begin{displaymath}
  \Pi(i\nu)
    = \frac{1}{2}
      \int\mathrm{d}\varepsilon\,
      \widehat{G}(\varepsilon)\,
      \int\mathrm{d}\varepsilon\,
      \widehat{G}(\varepsilon')\,
      \frac{f(\varepsilon - i\frac{\pi}{2\beta}) -
            f(\varepsilon' - i\frac{\pi}{2\beta})}
           {i\nu - \varepsilon + \varepsilon'}
\end{displaymath}
Note that the imaginary-valued chemical potential $\mu^f=
i\frac{\pi}{2\beta}$ cancels in the denominator, since $\Pi$
represents an observable susceptibility. Apparently, $\Pi(i\nu)$ obeys
the usual spectral representation, similar to
Eq.(\ref{eqn-deriv-effint}), with the imaginary part
\begin{eqnarray}
  \label{eqn-deriv-piimag}
  \Pi''(\omega)
    & = & \frac{\pi}{2}
      \int\mathrm{d}\varepsilon\,
      \widehat{G}(\varepsilon)
      \widehat{G}(\varepsilon - \omega)\,\cdot
      \\  \nonumber
    &   & \mbox{}\quad\cdot
      {}[\,f(\varepsilon - \omega - i\frac{\pi}{2\beta}) -
           f(\varepsilon - i\frac{\pi}{2\beta})\,{}]
\end{eqnarray}
and the corresponding real part $\Pi'$ is computed via
Eq.(\ref{eqn-nflex-pire})\,.

It is convenient to introduce a structure factor for the bubble,
\begin{displaymath}
  S^0(\omega)
    = {}[\,1 + g(\omega)\,{}]\,
      \Pi''(\omega)\;\;,
\end{displaymath}
and with the relation
\begin{displaymath}
  {}[\,f(x - y) - f(x)\,{}]
  {}[\,1 + g(y)\,{}]
    = {}[\,1 - f(x)\,{}]\,f(x - y)\;\;,
\end{displaymath}
which is valid for arbitrary complex numbers $x$\,, $y$\,, we have
\begin{displaymath}
  S^0(\omega)
    = \frac{\pi}{2}
      \int\mathrm{d}\varepsilon\,
      \widehat{G}(\varepsilon)
      \widehat{G}(\varepsilon - \omega)\,
      {}[\,1 - f(\varepsilon - i\frac{\pi}{2\beta})\,{}]\,
      f(\varepsilon - \omega - i\frac{\pi}{2\beta})
\end{displaymath}
Using the notation $\widehat{G}^+$\,, $\widehat{G}^-$ introduced
Eqs.(\ref{eqn-nflex-gfplus}), (\ref{eqn-nflex-gfmin}), the result
stated in Eq.(\ref{eqn-nflex-snul}) immediately follows.

In order to compute $\Pi''(\omega)$ from $S^0(\omega)$\,,
Eq.(\ref{eqn-nflex-piim}) is used, which is a consequence of the
symmetry $\Pi''(-\omega)= -\Pi''(\omega)$\,.

The last equation to be derived in this appendix is
Eq.(\ref{eqn-nflex-effint}) for the effective interaction
$U(\omega)$\,. Performing the analytic continuation $i\nu\to (\omega +
i0_+)$ in Eqs.(\ref{eqn-sflex-effint}), (\ref{eqn-sflex-chi}), and
using the decomposition (\ref{eqn-deriv-cont}), we find
\begin{displaymath}
  D''(\omega)
    = \frac{1}{N_L}
      \sum_{\mathbf{q}}
      J^2(\mathbf{q})
      \frac{\Pi''(\omega)}
           {\big|1 + J(\mathbf{q})\Pi(\omega + i0_+)\big|^2}\;.
\end{displaymath}
Eq.(\ref{eqn-nflex-effint}) is now obtained using the definition of
$U(\omega)$ and $S^0(\omega)$ given in this appendix and the
density-of-states $\mathcal{N}(\varepsilon)$ introduced in
Eq.(\ref{eqn-dos})\,.

\emph{Particle--hole symmetry:} In the Appendix~\ref{sec-app-greens}
above, a symmetry for the spectrum $\widehat{G}$ of the fermion
Green's function has been derived in Eq.(\ref{eqn-green-phsym}),
namely
\begin{displaymath}
  \widehat{G}(-\omega)
    = \widehat{G}(\omega)^\ast\;.
\end{displaymath}
Accordingly, the spectra $\widehat{G}^+$ and $\widehat{G}^-$
introduced in Eqs.(\ref{eqn-nflex-gfplus}), (\ref{eqn-nflex-gfmin})
obey the relation
\begin{equation}
  \label{eqn-simpl-phsymm}
  \widehat{G}^-(-\omega)
    = \widehat{G}^{+}(\omega)^\ast\;.
\end{equation}
This may be used to simplify the Eqs.(\ref{eqn-nflex}) somewhat by
eliminating $\widehat{G}^-$\,: Applying Eq.(\ref{eqn-simpl-phsymm}) to
Eq.(\ref{eqn-nflex-selfspec}) leads to
\begin{subequations}
\begin{equation}
  \label{eqn-simpl-selfspec}
  \widehat{\Sigma}(\omega)
    = \frac{3}{4\pi}
      \int\mathrm{d}\varepsilon\,
      U(\varepsilon)
      \,{}[\,
        \widehat{G}^+(\omega - \varepsilon) \;+\;
        \widehat{G}^{+}(-\omega - \varepsilon)^\ast
        \,{}]\;.
\end{equation}
For $S^0$\,, we start from Eq.(\ref{eqn-nflex-snul}) by writing the
expression twice and using the symmetry (\ref{eqn-simpl-phsymm}) in
the second term,
\begin{displaymath}
  S^0(\omega)
    = \frac{\pi}{4}
      \int\mathrm{d}\varepsilon\,
      {}[\,\widehat{G}^+(\varepsilon)
           \widehat{G}^-(\varepsilon - \omega) \;+\;
           \widehat{G}^{-}(-\varepsilon)^\ast
           \widehat{G}^{+}(\omega -\varepsilon)^\ast
           \,{}]
\end{displaymath}
By renamimg $\varepsilon\to(\omega - \varepsilon)$ in the second term
and applying Eq.(\ref{eqn-simpl-phsymm}) once more, it follows
\begin{equation}
  \label{eqn-simpl-snul}
  S^0(\omega)
    = \frac{\pi}{2}
      \int\mathrm{d}\varepsilon\,
      \textrm{Re}\big\{
        \widehat{G}^+(\varepsilon)\,
        \widehat{G}^{+}(\omega - \varepsilon)^\ast
        \big\}\;.
\end{equation}
\end{subequations}
The remaining equations in (\ref{eqn-nflex}) stay unchanged, except
that Eq.(\ref{eqn-nflex-gfmin}) becomes obsolete.

\section{The self-consistent equations using real spectra}
\label{sec-app-simpl}

For a direct comparison of the self-consistent equations with those
derived within average projection in Ref.\ \onlinecite{briwol04}
(Eqs.(A1) in that reference), we find it instructive to re-write the
Eqs.(\ref{eqn-nflex}) entirely in real-valued spectral functions.  To
that end we decompose all unphysical spectra into real and imaginary
parts as follows,
\begin{subequations}
  \label{eqn-simpl-decomp}
\begin{eqnarray}
  \widehat{G}(\omega)
    & = & \rho_1(\omega) + i\rho_2(\omega)\;\;,
    \\
  \widehat{G}^+(\omega)
    & = & \rho_1^+(\omega) + i\rho_2^+(\omega)\;\;,
    \\
  \widehat{\Sigma}(\omega)
    & = & \hat{\sigma}_1(\omega) + i\hat{\sigma}_2(\omega)\;\;,
    \\
  \overline{\Sigma}(\omega)
    & = & \bar{\sigma}_1(\omega) + i\bar{\sigma}_2(\omega)\;.
\end{eqnarray}
\end{subequations}
Inserting these definitions in Eqs.(\ref{eqn-nflex}), with
Eqs.(\ref{eqn-nflex-selfspec}) and (\ref{eqn-nflex-snul}) replaced by
Eqs.(\ref{eqn-simpl-selfspec}) and (\ref{eqn-simpl-snul}), we find the
set of equations (\ref{eqn-simpl-num}) stated below. For completeness,
in (\ref{eqn-simpl-num}) we also quote those equations from
(\ref{eqn-nflex}), that remain unchanged by using
(\ref{eqn-simpl-decomp})\,. Making use of Eq.(\ref{eqn-fermicmplx})
the result reads
\begin{subequations}
  \label{eqn-simpl-num}
\begin{eqnarray}
  U(\omega)
    & = & S^0(\omega)
      \int\mathrm{d}\varepsilon\,
      \frac{\mathcal{N}(\varepsilon)\,\varepsilon^2}
           {\big|1 + \varepsilon\,\Pi(\omega + i0_+)\big|^2}
      \\
  \Pi''(\omega)
    & = & S^0(\omega) - S^0(-\omega)
      \\
  \Pi'(\omega)
    & = & \frac{1}{\pi}
      \textrm{P}\!\!\int\mathrm{d}\varepsilon\,
      \frac{\Pi''(\varepsilon)}{\varepsilon - \omega}
      \\
  S^0(\omega)
    & = & \frac{\pi}{2}
      \int\mathrm{d}\varepsilon\,
      {}\big[\,\rho_1^+(\varepsilon) \rho_1^+(\omega - \varepsilon)
      + \\
    &   & \nonumber
      \mbox{}\quad\quad
      + \rho_2^+(\varepsilon) \rho_2^+(\omega - \varepsilon)
      \,{}\big]
      \\
  \rho_1^+(\omega)
    & = & {}[\,1 - f(2\omega)\,{}]\,\rho_1(\omega) \;+\;
      \frac{\rho_2(\omega)}{2\cosh(\beta\omega)}
      \\
  \rho_2^+(\omega)
    & = & {}[\,1 - f(2\omega)\,{}]\,\rho_2(\omega) \;-\;
      \frac{\rho_1(\omega)}{2\cosh(\beta\omega)}
      \\
  \rho_1(\omega)
    & = & \\ \nonumber
  \lefteqn{\textrm{Re}
      \frac{\hat{\sigma}_1(\omega) + i\hat{\sigma}_2(\omega)}
           {(A_+ A_- - B_+ B_-) - i(B_+ A_- + A_+ B_-)}}
      \\
  \rho_2(\omega)
    & = & \\ \nonumber
  \lefteqn{\textrm{Im}
      \frac{\hat{\sigma}_1(\omega) + i\hat{\sigma}_2(\omega)}
           {(A_+ A_- - B_+ B_-) - i(B_+ A_- + A_+ B_-)}}
      \\
  \hat{\sigma}_1(\omega)
    & = & \frac{3}{4\pi}\int\mathrm{d}\varepsilon\,U(\varepsilon)
      {}[\,\rho_1^+(\omega - \varepsilon) +
      \\ \nonumber
    &   & \mbox{}\quad\quad
      + \rho_1^+(-\omega - \varepsilon)\,{}]
      \\
  \hat{\sigma}_2(\omega)
    & = & \frac{3}{4\pi}\int\mathrm{d}\varepsilon\,U(\varepsilon)
      {}[\,\rho_2^+(\omega - \varepsilon) -
      \\ \nonumber
    &   & \mbox{}\quad\quad
      - \rho_2^+(-\omega - \varepsilon)\,{}]
      \\
  \bar{\sigma}_1(\omega)
    & = & \textrm{P}\!\!\int\mathrm{d}\varepsilon\,
      \frac{\hat{\sigma}_1(\varepsilon)}{\omega - \varepsilon}
      \\
  \bar{\sigma}_2(\omega)
    & = & \textrm{P}\!\!\int\mathrm{d}\varepsilon\,
      \frac{\hat{\sigma}_2(\varepsilon)}{\omega - \varepsilon}
\end{eqnarray}
\end{subequations}
The short hands $A_{\pm}$\,, $B_{\pm}$ are defined as
\begin{eqnarray*}
  A_{\pm}
    & = & \omega - \bar{\sigma}_1(\omega)
      \pm \pi\hat{\sigma}_2(\omega)
      \\
  B_{\pm}
    & = & \bar{\sigma}_2(\omega) \pm \pi\hat{\sigma}_1(\omega)
\end{eqnarray*}

\section{Calculation of the auxiliary-charge response
  (\ref{eqn-flex-qbub})}
\label{sec-app-qbub}

In this appendix the intermediate steps in going from
Eq.(\ref{eqn-flex-bethe}) to Eq.(\ref{eqn-flex-bethe2}) are
explained. We start by writing Eq.(\ref{eqn-flex-bethe}) in the form
\begin{eqnarray}
  \label{eqn-qbub-bethe}
  \Gamma(\mathbf{q}, i\omega)
    & = & 1 + \sum_{i\omega_1}
      \big[\,A(i\omega, i\omega_1) +
      \\  \nonumber
    &   & \quad\mbox{} +
      B(\mathbf{q}; i\omega, i\omega_1)\,\big]
      \,\Gamma(\mathbf{q}, i\omega_1)
\end{eqnarray}
with the matrices
\begin{displaymath}
  A(i\omega, i\omega_1)
    = \frac{3}{4}T\,G(i\omega_1)^2 D(i\omega - i\omega_1)
\end{displaymath}
and $B(\mathbf{q}; i\omega, i\omega_1)$\,, the latter can be read off
Eq.(\ref{eqn-flex-bethe})\,. For the constituents of $A$ and $B$ we
observe the symmetries
\begin{displaymath}
  G(-i\omega)= -G(i\omega)  \;\;,\;\;\;
  D(\mathbf{k}, -i\nu)
    = D(\mathbf{k}, i\nu) \;\;,\;\;\; 
\end{displaymath}
which implies
\begin{subequations}
  \label{eqn-qbub-symm}
\begin{equation}
  A(-i\omega, -i\omega_1)
    = A(i\omega, i\omega_1)
\end{equation}
and
\begin{equation}
\begin{array}[c]{rcl}
  B(-i\omega, i\omega_1)
    & = & - B(i\omega, i\omega_1)\;\;,
      \\[1ex]
  B(i\omega, -i\omega_1)
    & = & - B(i\omega, i\omega_1)\;.
\end{array}
\end{equation}
\end{subequations}
Here and in the following the wave vector $\mathbf{q}$ is not written.
The vertex function is split into components behaving symmetrical
$(+)$ or anti-symmetrical $(-)$ under $i\omega\to -i\omega$\,,
\begin{displaymath}
  \Gamma(i\omega)
    = \Gamma^{(+)}(i\omega) + \Gamma^{(-)}(i\omega)\;\;,
\end{displaymath}
corresponding to
\begin{displaymath}
  \Gamma^{(\pm)}(i\omega)
    = \frac{1}{2}{}[\,\Gamma(i\omega) \pm \Gamma(-i\omega)\,{}]\;.
\end{displaymath}
Using these definitions and the symmetry relations
(\ref{eqn-qbub-symm}) in Eq.(\ref{eqn-qbub-bethe}), it can be seen
that $\Gamma^{(+)}$ and $\Gamma^{(-)}$ decouple,
\begin{subequations}
  \label{eqn-qbub-split}
\begin{equation}
  \label{eqn-qbub-splitplus}
  \Gamma^{(+)}(i\omega)
    = 1 + \sum_{i\omega_1}
      A(i\omega, i\omega_1)\,
      \Gamma^{(+)}(i\omega_1)\;\;,
\end{equation}
\begin{equation}
  \Gamma^{(-)}(i\omega)
    = \sum_{i\omega_1}
      \big[\,A(i\omega, i\omega_1) +
             B(i\omega, i\omega_1)\,\big]\,
      \Gamma^{(-)}(i\omega_1)\;.
\end{equation}
\end{subequations}
In the fermion bubble (\ref{eqn-flex-qbub}) only the symmetric
component contributes (not writing $\mathbf{q}$),
\begin{displaymath}
  \chi^Q(0)
    = -2T\sum_{i\omega}G(i\omega)^2\,
      \Gamma^{(+)}(i\omega)\;.
\end{displaymath}
Therefore, by identifying $\Gamma^{(+)}$ with $\Gamma$ in
Eq.(\ref{eqn-qbub-splitplus}), we arrive at Eq.(\ref{eqn-flex-bethe2})
for the vertex function to be used in the fermion-charge response
(\ref{eqn-flex-qbub})\,.


\begin{thebibliography}{40}
\expandafter\ifx\csname natexlab\endcsname\relax\def\natexlab#1{#1}\fi
\expandafter\ifx\csname bibnamefont\endcsname\relax
  \def\bibnamefont#1{#1}\fi
\expandafter\ifx\csname bibfnamefont\endcsname\relax
  \def\bibfnamefont#1{#1}\fi
\expandafter\ifx\csname citenamefont\endcsname\relax
  \def\citenamefont#1{#1}\fi
\expandafter\ifx\csname url\endcsname\relax
  \def\url#1{\texttt{#1}}\fi
\expandafter\ifx\csname urlprefix\endcsname\relax\def\urlprefix{URL }\fi
\providecommand{\bibinfo}[2]{#2}
\providecommand{\eprint}[2][]{\url{#2}}

\bibitem[{\citenamefont{Manousakis}(1991)}]{man91}
\bibinfo{author}{\bibfnamefont{E.}~\bibnamefont{Manousakis}},
  \bibinfo{journal}{Rev.~Mod.~Phys.~} \textbf{\bibinfo{volume}{63}},
  \bibinfo{pages}{1} (\bibinfo{year}{1991}).

\bibitem[{\citenamefont{Gros et~al.}(1987)\citenamefont{Gros, Joynt, and
  Rice}}]{grosjoyntrice87}
\bibinfo{author}{\bibfnamefont{C.}~\bibnamefont{Gros}},
  \bibinfo{author}{\bibfnamefont{R.}~\bibnamefont{Joynt}}, \bibnamefont{and}
  \bibinfo{author}{\bibfnamefont{T.~M.} \bibnamefont{Rice}},
  \bibinfo{journal}{Phys.~Rev.~B} \textbf{\bibinfo{volume}{36}},
  \bibinfo{pages}{381} (\bibinfo{year}{1987}).

\bibitem[{\citenamefont{Hayden et~al.}(1990)\citenamefont{Hayden, Aeppli, Mook,
  Cheong, and Fisk}}]{hay90}
\bibinfo{author}{\bibfnamefont{S.~M.} \bibnamefont{Hayden}},
  \bibinfo{author}{\bibfnamefont{G.}~\bibnamefont{Aeppli}},
  \bibinfo{author}{\bibfnamefont{H.~A.} \bibnamefont{Mook}},
  \bibinfo{author}{\bibfnamefont{S.-W.} \bibnamefont{Cheong}},
  \bibnamefont{and} \bibinfo{author}{\bibfnamefont{Z.}~\bibnamefont{Fisk}},
  \bibinfo{journal}{Phys.~Rev.~B} \textbf{\bibinfo{volume}{42}},
  \bibinfo{pages}{10220} (\bibinfo{year}{1990}).

\bibitem[{\citenamefont{Keimer et~al.}(1992)\citenamefont{Keimer, Belk,
  Birgeneau, Cassaho, Chen, Greven, Kastner, Aharony, Endoh, Erwin
  et~al.}}]{keim92}
\bibinfo{author}{\bibfnamefont{B.}~\bibnamefont{Keimer}},
  \bibinfo{author}{\bibfnamefont{N.}~\bibnamefont{Belk}},
  \bibinfo{author}{\bibfnamefont{R.~J.} \bibnamefont{Birgeneau}},
  \bibinfo{author}{\bibfnamefont{A.}~\bibnamefont{Cassaho}},
  \bibinfo{author}{\bibfnamefont{C.~Y.} \bibnamefont{Chen}},
  \bibinfo{author}{\bibfnamefont{M.}~\bibnamefont{Greven}},
  \bibinfo{author}{\bibfnamefont{M.~A.} \bibnamefont{Kastner}},
  \bibinfo{author}{\bibfnamefont{A.}~\bibnamefont{Aharony}},
  \bibinfo{author}{\bibfnamefont{Y.}~\bibnamefont{Endoh}},
  \bibinfo{author}{\bibfnamefont{R.~W.} \bibnamefont{Erwin}},
  \bibnamefont{et~al.}, \bibinfo{journal}{Phys.~Rev.~B}
  \textbf{\bibinfo{volume}{46}}, \bibinfo{pages}{14034} (\bibinfo{year}{1992}).

\bibitem[{\citenamefont{Kim et~al.}(2001)\citenamefont{Kim, Birgeneau, Chou,
  Erwin, and Kastner}}]{kim01}
\bibinfo{author}{\bibfnamefont{Y.~J.} \bibnamefont{Kim}},
  \bibinfo{author}{\bibfnamefont{R.~J.} \bibnamefont{Birgeneau}},
  \bibinfo{author}{\bibfnamefont{F.~C.} \bibnamefont{Chou}},
  \bibinfo{author}{\bibfnamefont{R.~W.} \bibnamefont{Erwin}}, \bibnamefont{and}
  \bibinfo{author}{\bibfnamefont{M.~A.} \bibnamefont{Kastner}},
  \bibinfo{journal}{Phys.~Rev.~Lett.~} \textbf{\bibinfo{volume}{86}},
  \bibinfo{pages}{3144} (\bibinfo{year}{2001}).

\bibitem[{\citenamefont{Negele and Orland}(1988)}]{negorl}
\bibinfo{author}{\bibfnamefont{J.~W.} \bibnamefont{Negele}} \bibnamefont{and}
  \bibinfo{author}{\bibfnamefont{H.}~\bibnamefont{Orland}},
  \emph{\bibinfo{title}{Quantum Many-Particle Systems}}
  (\bibinfo{publisher}{Addison-Wesley}, \bibinfo{address}{Menlo Part etc.~},
  \bibinfo{year}{1988}).

\bibitem[{\citenamefont{Grewe and Keiter}(1981)}]{grekei81}
\bibinfo{author}{\bibfnamefont{N.}~\bibnamefont{Grewe}} \bibnamefont{and}
  \bibinfo{author}{\bibfnamefont{H.}~\bibnamefont{Keiter}},
  \bibinfo{journal}{Phys.~Rev.~B} \textbf{\bibinfo{volume}{24}},
  \bibinfo{pages}{4420} (\bibinfo{year}{1981}).

\bibitem[{\citenamefont{Bickers}(1987)}]{bic87}
\bibinfo{author}{\bibfnamefont{N.~E.} \bibnamefont{Bickers}},
  \bibinfo{journal}{Rev.~Mod.~Phys.~} \textbf{\bibinfo{volume}{59}},
  \bibinfo{pages}{846} (\bibinfo{year}{1987}).

\bibitem[{\citenamefont{Izyumov and Skryabin}(1988)}]{russenbuch}
\bibinfo{author}{\bibfnamefont{Y.~A.} \bibnamefont{Izyumov}} \bibnamefont{and}
  \bibinfo{author}{\bibfnamefont{Y.~N.} \bibnamefont{Skryabin}},
  \emph{\bibinfo{title}{Statistical Mechanics of Magnetically Ordered Systems}}
  (\bibinfo{publisher}{Consultants Bureau}, \bibinfo{address}{New York},
  \bibinfo{year}{1988}).

\bibitem[{\citenamefont{Abrikosov}(1965)}]{abr65}
\bibinfo{author}{\bibfnamefont{A.~A.} \bibnamefont{Abrikosov}},
  \bibinfo{journal}{Physics} \textbf{\bibinfo{volume}{2}}, \bibinfo{pages}{5}
  (\bibinfo{year}{1965}).

\bibitem[{\citenamefont{Arovas and Auerbach}(1988)}]{aaa88}
\bibinfo{author}{\bibfnamefont{D.~P.} \bibnamefont{Arovas}} \bibnamefont{and}
  \bibinfo{author}{\bibfnamefont{A.}~\bibnamefont{Auerbach}},
  \bibinfo{journal}{Phys.~Rev.~B} \textbf{\bibinfo{volume}{38}},
  \bibinfo{pages}{316} (\bibinfo{year}{1988}).

\bibitem[{\citenamefont{Abrikosov and Migdal}(1970)}]{abrmig70}
\bibinfo{author}{\bibfnamefont{A.~A.} \bibnamefont{Abrikosov}}
  \bibnamefont{and} \bibinfo{author}{\bibfnamefont{A.~A.}
  \bibnamefont{Migdal}}, \bibinfo{journal}{J.~Low~Temp.~Phys.~}
  \textbf{\bibinfo{volume}{3}}, \bibinfo{pages}{519} (\bibinfo{year}{1970}).

\bibitem[{\citenamefont{Bickers et~al.}(1987)\citenamefont{Bickers, Cox, and
  Wilkins}}]{biccoxwil87}
\bibinfo{author}{\bibfnamefont{N.~E.} \bibnamefont{Bickers}},
  \bibinfo{author}{\bibfnamefont{D.~L.} \bibnamefont{Cox}}, \bibnamefont{and}
  \bibinfo{author}{\bibfnamefont{J.~W.} \bibnamefont{Wilkins}},
  \bibinfo{journal}{Phys.~Rev.~B} \textbf{\bibinfo{volume}{36}},
  \bibinfo{pages}{2036} (\bibinfo{year}{1987}).

\bibitem[{\citenamefont{Coleman}(1984)}]{col84}
\bibinfo{author}{\bibfnamefont{P.}~\bibnamefont{Coleman}},
  \bibinfo{journal}{Phys.~Rev.~B} \textbf{\bibinfo{volume}{29}},
  \bibinfo{pages}{3035} (\bibinfo{year}{1984}).

\bibitem[{\citenamefont{Brout}(1961)}]{bro61}
\bibinfo{author}{\bibfnamefont{R.}~\bibnamefont{Brout}},
  \bibinfo{journal}{Phys.~Rev.~} \textbf{\bibinfo{volume}{122}},
  \bibinfo{pages}{469} (\bibinfo{year}{1961}).

\bibitem[{\citenamefont{Kuramoto}(1985)}]{kur85}
\bibinfo{author}{\bibfnamefont{Y.}~\bibnamefont{Kuramoto}}, in
  \emph{\bibinfo{booktitle}{{Theory of Heavy Fermions {and} Valence
  Fluctuations}}}, edited by
  \bibinfo{editor}{\bibfnamefont{T.}~\bibnamefont{Kasuya}} \bibnamefont{and}
  \bibinfo{editor}{\bibfnamefont{T.}~\bibnamefont{Saso}}
  (\bibinfo{publisher}{Springer}, \bibinfo{address}{Berlin},
  \bibinfo{year}{1985}), p. \bibinfo{pages}{152}.

\bibitem[{\citenamefont{Metzner and Vollhardt}(1989)}]{metvol89}
\bibinfo{author}{\bibfnamefont{W.}~\bibnamefont{Metzner}} \bibnamefont{and}
  \bibinfo{author}{\bibfnamefont{D.}~\bibnamefont{Vollhardt}},
  \bibinfo{journal}{Phys.~Rev.~Lett.~} \textbf{\bibinfo{volume}{62}},
  \bibinfo{pages}{324} (\bibinfo{year}{1989}).

\bibitem[{\citenamefont{Si and Smith}(1996)}]{sismith96}
\bibinfo{author}{\bibfnamefont{Q.}~\bibnamefont{Si}} \bibnamefont{and}
  \bibinfo{author}{\bibfnamefont{J.~L.} \bibnamefont{Smith}},
  \bibinfo{journal}{Phys.~Rev.~Lett.~} \textbf{\bibinfo{volume}{77}},
  \bibinfo{pages}{3391} (\bibinfo{year}{1996}).

\bibitem[{not({\natexlab{a}})}]{note-dca}
\bibinfo{note}{See, e.g., Ref.\ \onlinecite{kot01} and the references given
  therein.}

\bibitem[{\citenamefont{Newns and Read}(1987)}]{newrea87}
\bibinfo{author}{\bibfnamefont{D.~M.} \bibnamefont{Newns}} \bibnamefont{and}
  \bibinfo{author}{\bibfnamefont{N.}~\bibnamefont{Read}},
  \bibinfo{journal}{Adv.~Physics} \textbf{\bibinfo{volume}{36}},
  \bibinfo{pages}{799} (\bibinfo{year}{1987}).

\bibitem[{\citenamefont{Popov and Fedotov}(1988)}]{popfed88}
\bibinfo{author}{\bibfnamefont{V.~N.} \bibnamefont{Popov}} \bibnamefont{and}
  \bibinfo{author}{\bibfnamefont{S.~A.} \bibnamefont{Fedotov}},
  \bibinfo{journal}{Sov.~Phys.~JETP} \textbf{\bibinfo{volume}{67}},
  \bibinfo{pages}{535} (\bibinfo{year}{1988}).

\bibitem[{\citenamefont{Gros and Johnson}(1990)}]{grosjohn90}
\bibinfo{author}{\bibfnamefont{C.}~\bibnamefont{Gros}} \bibnamefont{and}
  \bibinfo{author}{\bibfnamefont{M.~D.} \bibnamefont{Johnson}},
  \bibinfo{journal}{Physica B} \textbf{\bibinfo{volume}{165\,\&\,166}},
  \bibinfo{pages}{985} (\bibinfo{year}{1990}).

\bibitem[{\citenamefont{Bouis and Kiselev}(1999)}]{boukis99}
\bibinfo{author}{\bibfnamefont{F.}~\bibnamefont{Bouis}} \bibnamefont{and}
  \bibinfo{author}{\bibfnamefont{M.~N.} \bibnamefont{Kiselev}},
  \bibinfo{journal}{Physica B} \textbf{\bibinfo{volume}{259--261}},
  \bibinfo{pages}{195} (\bibinfo{year}{1999}).

\bibitem[{\citenamefont{Kiselev and Oppermann}(2000)}]{kisopp00}
\bibinfo{author}{\bibfnamefont{M.~N.} \bibnamefont{Kiselev}} \bibnamefont{and}
  \bibinfo{author}{\bibfnamefont{R.}~\bibnamefont{Oppermann}},
  \bibinfo{journal}{JETP Lett.} \textbf{\bibinfo{volume}{71}},
  \bibinfo{pages}{250} (\bibinfo{year}{2000}).

\bibitem[{\citenamefont{Dillenschneider and Richert}(2006)}]{dilric06}
\bibinfo{author}{\bibfnamefont{R.}~\bibnamefont{Dillenschneider}}
  \bibnamefont{and} \bibinfo{author}{\bibfnamefont{J.}~\bibnamefont{Richert}},
  \bibinfo{journal}{Phys.~Rev.~B} \textbf{\bibinfo{volume}{73}},
  \bibinfo{pages}{024409} (\bibinfo{year}{2006}).

\bibitem[{\citenamefont{Luttinger and Ward}(1960)}]{lutwar60}
\bibinfo{author}{\bibfnamefont{J.~M.} \bibnamefont{Luttinger}}
  \bibnamefont{and} \bibinfo{author}{\bibfnamefont{J.~C.} \bibnamefont{Ward}},
  \bibinfo{journal}{Phys.~Rev.~} \textbf{\bibinfo{volume}{118}},
  \bibinfo{pages}{1417} (\bibinfo{year}{1960}).

\bibitem[{\citenamefont{Baym}(1962)}]{bay62}
\bibinfo{author}{\bibfnamefont{G.}~\bibnamefont{Baym}},
  \bibinfo{journal}{Phys.~Rev.~} \textbf{\bibinfo{volume}{127}},
  \bibinfo{pages}{1391} (\bibinfo{year}{1962}).

\bibitem[{not({\natexlab{b}})}]{note-oldref}
\bibinfo{note}{A number of references has been given in Ref.\
  \onlinecite{briwol04}.}

\bibitem[{\citenamefont{Brinckmann and W{\"o}lfle}(2004)}]{briwol04}
\bibinfo{author}{\bibfnamefont{J.}~\bibnamefont{Brinckmann}} \bibnamefont{and}
  \bibinfo{author}{\bibfnamefont{P.}~\bibnamefont{W{\"o}lfle}},
  \bibinfo{journal}{Phys.~Rev.~B} \textbf{\bibinfo{volume}{70}},
  \bibinfo{pages}{174445} (\bibinfo{year}{2004}).

\bibitem[{\citenamefont{Bickers et~al.}(1989)\citenamefont{Bickers, Scalapino,
  and White}}]{bicscawhi89}
\bibinfo{author}{\bibfnamefont{N.~E.} \bibnamefont{Bickers}},
  \bibinfo{author}{\bibfnamefont{D.~J.} \bibnamefont{Scalapino}},
  \bibnamefont{and} \bibinfo{author}{\bibfnamefont{S.~R.} \bibnamefont{White}},
  \bibinfo{journal}{Phys.~Rev.~Lett.~} \textbf{\bibinfo{volume}{62}},
  \bibinfo{pages}{961} (\bibinfo{year}{1989}).

\bibitem[{foo()}]{foot-cons}
\bibinfo{note}{The term ``measurable'' referes to the fact that $\chi$ is a
  correlation function of physical, i.e., gauge-invariant operators, in
  contrast to the fermion Green's function. There is the separate issue,
  discussed in Ref.\ \onlinecite{briwol04}\,, whether the susceptibility that
  is compared to experiment should be calculated with additional vertex
  corrections.}

\bibitem[{not({\natexlab{c}})}]{note-num}
\bibinfo{note}{In Ref.\ \onlinecite{briwol04} we reported $a= 0.30$ and $b=
  0.0$ for the case of average projection. While the value of $a$ is reproduced
  here, $b\simeq 0.5$ comes out different. This is due to an improved numerical
  iteration scheme and accuracy used in the present work.}

\bibitem[{\citenamefont{Chakravarty et~al.}(1989)\citenamefont{Chakravarty,
  Halperin, and Nelson}}]{chn89}
\bibinfo{author}{\bibfnamefont{S.}~\bibnamefont{Chakravarty}},
  \bibinfo{author}{\bibfnamefont{B.~I.} \bibnamefont{Halperin}},
  \bibnamefont{and} \bibinfo{author}{\bibfnamefont{D.~R.}
  \bibnamefont{Nelson}}, \bibinfo{journal}{Phys.~Rev.~B}
  \textbf{\bibinfo{volume}{39}}, \bibinfo{pages}{2344} (\bibinfo{year}{1989}).

\bibitem[{not({\natexlab{d}})}]{note-fspecen}
\bibinfo{note}{This can be seen from a close inspection of the numerical data
  Fig.\ \ref{fig-fdos} is made of. The numerical observation is supported by an
  analytical solution of the Eqs.(\ref{eqn-nflex}) using a ``static''
  approximation similar to the one introduced in Ref.\
  \onlinecite{briwol04}\,.}

\bibitem[{not({\natexlab{e}})}]{note-qfluct-hartree}
\bibinfo{note}{There is no vertex function in the charge channel in Hartree
  approximation, since there is no direct charge--charge interaction.}

\bibitem[{not({\natexlab{f}})}]{note-pregap}
\bibinfo{note}{As has been discussed in Ref.\ \onlinecite{briwol04}\,, the
  self-consistent approximation shown in Fig.\ \ref{fig-flex} lacks the
  formation of a ``precursor gap'', which is expected to appear in the fermion
  spectrum when the ordered ground state is approached. In Ref.\
  \onlinecite{briwol04} we have proposed the ``MSCA'' as an alternative
  approximation, which does feature a precursor-pseudo gap. However, the latter
  cannot be used here, since we don't have a $\Phi$-functional (conserving
  approximation) and therefore no consistent expression for $\langle \Delta
  Q_i\,\Delta Q_j\rangle$\,.}

\bibitem[{not({\natexlab{g}})}]{note-qtransfer}
\bibinfo{note}{Strictly spoken, the total fermion charge $\sum_{i= 1}^{N_L}Q_i$
  is conserved, even in average projection, since $N_L$ is macroscopically
  large. Therefore the simplest $Q$-fluctuation is a transfer of a fermion from
  one site, say, $l$ to another site $m$\,, equivalent to the removal of the
  two spins at sites $l$ and $m$\,.}

\bibitem[{not({\natexlab{h}})}]{note-qflexpon}
\bibinfo{note}{From that argument, the charge fluctuations should always show
  activated behavior, $\langle \Delta Q_i\,\Delta Q_j\rangle\sim\exp(-J/T)$\,,
  but see the discussion below Eq.(\ref{eqn-qflu-flex}) and footnote
  \onlinecite{note-pregap}\,.}

\bibitem[{\citenamefont{Misguich and Lhuillier}(2005)}]{mislhu05}
\bibinfo{author}{\bibfnamefont{G.}~\bibnamefont{Misguich}} \bibnamefont{and}
  \bibinfo{author}{\bibfnamefont{C.}~\bibnamefont{Lhuillier}}, in
  \emph{\bibinfo{booktitle}{Frustrated spin systems}}, edited by
  \bibinfo{editor}{\bibfnamefont{H.~T.} \bibnamefont{Diep}}
  (\bibinfo{publisher}{World-Scientific}, \bibinfo{year}{2005}),
  \bibinfo{note}{{ISBN 981-256-091-2}}.

\bibitem[{\citenamefont{Kotliar et~al.}(2001)\citenamefont{Kotliar, Savrasov,
  P{\'a}lsson, and Biroli}}]{kot01}
\bibinfo{author}{\bibfnamefont{G.}~\bibnamefont{Kotliar}},
  \bibinfo{author}{\bibfnamefont{S.~Y.} \bibnamefont{Savrasov}},
  \bibinfo{author}{\bibfnamefont{G.}~\bibnamefont{P{\'a}lsson}},
  \bibnamefont{and} \bibinfo{author}{\bibfnamefont{G.}~\bibnamefont{Biroli}},
  \bibinfo{journal}{Phys.~Rev.~Lett.~} \textbf{\bibinfo{volume}{87}},
  \bibinfo{pages}{186401} (\bibinfo{year}{2001}).

\end{thebibliography}
\end{document}